\documentclass[journal=apchd5,manuscript=article,layout=twocolumn]{achemso}%

\makeatletter
\let\l@addto@macro\relax
\makeatother
\usepackage[fontsize=10pt]{scrextend}

\usepackage{geometry}
\usepackage{listings}
\usepackage[normalem]{ulem}
\usepackage{amsmath}
\usepackage{amsfonts}
\usepackage{amssymb}
\usepackage{hyperref} 
\usepackage{color}

\newcommand{\Bolivarallee}{Boliva\hspace{-0.1mm}r\hspace{0.15mm}a\hspace{-0.1mm}llee}
\newcommand{\Abbestrasse}{Abbe\hspace{0.15mm}s\hspace{-0.1mm}tra{\ss}e}
\newcommand{\Takustrasse}{Taku\hspace{0.15mm}s\hspace{-0.1mm}tra{\ss}e}

\newcommand{\AlGaAs}{Al$_{0.9}$Ga$_{0.1}$As}

\def\added#1{#1} 
\def\removed#1{} 

\author{Philipp-Immanuel~Schneider}
\affiliation[JCMwave GmbH]{JCMwave GmbH, \Bolivarallee~22, D\,--\,14\,050 Berlin, Germany}
\email{philipp.schneider@jcmwave.com}
\author{Xavier~Garcia~Santiago}
\affiliation[JCMwave GmbH]{JCMwave GmbH, \Bolivarallee~22, D\,--\,14\,050 Berlin, Germany}
\alsoaffiliation[KIT]{Institut f\"ur Nanotechnology, Karlsruher Institut f\"ur Technologie, PO-Box 3640, D\,--\,76\,021 Karlsruhe, Germany}
\author{Victor~Soltwisch}
\affiliation[PTB]{Physikalisch-Technische Bundesanstalt\,(PTB), \Abbestrasse~ 2-12, D\,--\,10\,587 Berlin, Germany}
\author{Martin~Hammerschmidt}
\affiliation[JCMwave GmbH]{JCMwave GmbH, \Bolivarallee~22, D\,--\,14\,050 Berlin, Germany}
\author{Sven~Burger}
\affiliation[JCMwave GmbH]{JCMwave GmbH, \Bolivarallee~22, D\,--\,14\,050 Berlin, Germany}
\alsoaffiliation[ZIB]
{Zuse Institute Berlin\,(ZIB), \Takustrasse~7, D\,--\,14\,195 Berlin, Germany}
\author{Carsten~Rockstuhl}
\affiliation[KIT]{Institut f\"ur Nanotechnology, Karlsruher Institut f\"ur Technologie, PO-Box 3640, D\,--\,76\,021 Karlsruhe, Germany}
\alsoaffiliation[KIT]
{Institut f\"ur Theoretische Festk\"orperphysik, Karlsruher Institut f\"ur Technologie, Wolfgang-Gaede-Str.~1, D\,--\,76\,131 Karlsruhe, Germany}

\title{Benchmarking five global optimization approaches for nano-optical shape optimization and parameter reconstruction}

\keywords{Shape optimization, Parameter reconstruction, Machine learning, Global optimization, Bayesian optimziation}

\begin{document}
This document is the unedited Author's version of a Submitted Work that was subsequently accepted for publication in ACS Photonics, copyright \textcopyright American Chemical Society after peer review. To access the final edited and published work see \href{https://doi.org/10.1021/acsphotonics.9b00706}{https://doi.org/10.1021/acsphotonics.9b00706}.

\begin{abstract}
  Numerical optimization is an important tool in the field of computational physics in general and in nano-optics in specific. It has attracted attention with the increase in complexity of structures that can be realized with nowadays nano-fabrication technologies for which a rational design is no longer feasible. Also, numerical resources are available to enable the computational photonic material design and to identify structures that meet predefined optical properties for specific applications. However, the optimization objective function is in general non-convex and its computation remains resource demanding such that the right choice for the optimization method is crucial to obtain excellent results. Here, we benchmark five global optimization methods for three typical nano-optical optimization problems: \removed{downhill simplex optimization, the limited-memory Broyden-Fletcher-Goldfarb-Shanno (L-BFGS) algorithm, particle swarm optimization, differential evolution, and Bayesian optimization} \added{particle swarm optimization, differential evolution, and Bayesian optimization as well as multi-start versions of downhill simplex optimization and the limited-memory Broyden-Fletcher-Goldfarb-Shanno (L-BFGS) algorithm}. In the shown examples from the field of shape optimization and parameter reconstruction, Bayesian optimization, mainly known from machine learning applications, obtains significantly better results in a fraction of the run times of the other optimization methods.
\end{abstract}

\phantom{1}

Numerical optimization is a fundamental task for many scientific and industrial applications. It is also an important tool in the field of nano-optics. Modern nano-processing technologies such as laser writing~\cite{doi:10.1002/lpor.201100046} or 3D in-situ electron-beam lithography~\cite{yoon2017fabrication,KAGANSKIY2018162} enable the fabrication of micro- and nano-optical structures with an increasing degree of accuracy and flexibility. From an experimental and technological perspective it is often not clear, which geometries and geometrical parameters lead to optimal results in terms of a desired functionality. Numerical simulation and scans of selected parameters can give important insights~\cite{Gschreyetal.2015,PhysRevB.81.125431,doi:10.1021/nl501648f,doi:10.1021/acsphotonics.5b00559}. However, the full exploitation of the fabrication flexibility requires the simultaneous numerical optimization of all degrees of freedom. This process can be very time consuming and prompts for large computing resources (e.~g. multi-core computers or computer clusters), fast simulation methods, and efficient numerical optimization methods that require as few as possible simulation runs of the actual forward problem~\cite{doi:10.1002/lpor.201700237}.

Another important application for numerical optimization is the parameter reconstruction based on measured data~\cite{pang2012computational}. For example, optical scatterometry is the state-of-the-art optical inspection technique for quality control in lithographic processing~\cite{bodermann2016optical}. This indirect measurement procedure relies on a parametrization of the specimen's geometry and a numerical simulation of the measurement process. Based on multiple numerical simulations, one tries to identify the parameters that match best the measured data. Especially, for in-line quality control it is crucial to find the parameters with as few simulation runs as possible. 

In each optimization scenario the first step is to define an objective function that maps the system's parameters to an objective value that is to be minimized. In nano-optics the computation of the objective function generally requires to solve Maxwell's equations. This can be achieved by different numerical methods depending on the geometry, such as the Finite-Element Method (FEM), rigorous coupled wave analysis (RCWA), and Finite-Difference Time-Domain (FDTD) method. In this work we use the software package JCMsuite~\cite{JCMsuite}, which employs the FEM approach in the frequency domain \cite{monk2003finite}. Typical computation times range from a few seconds for simple and highly symmetric systems to hours or even days for complex three-dimensional geometries with a spatial extent larger than many wavelengths. Nano-optical systems are often characterized by interference and resonance phenomena. Typically, by varying the dimensions of the system or the wavelength of the light, multiple resonances can be observed. As a consequence, the objective function features in general many separated minima, which makes it difficult to find the global minimum. This is, for example, in contrast to the optimization problem of training artificial neural networks, where local minima seem not to be an obstacle in finding optimal network weights~\cite{goodfellow2014qualitatively}. 

Optimization problems can be roughly divided into low-dimensional problems (1 to 3 parameters), medium-dimensional problems (4 to $\sim$ 15 parameters) and high-dimensional problems ($\sim$ 15 parameters to some hundred parameters or more). While low-dimensional problems often allow for scanning the full parameter space, this is already impossible for medium-dimensional problems. E.g., a scan of a 10-dimensional parameter space with a resolution of 100 different values for each parameter requires $100^{10}=10^{20}$ evaluations of the objective function, rendering a regular parameter scan infeasible. This problem, known as {\it curse of dimensionality}, is tackled by global optimization approaches, which try to sample the parameter space in an effective way by avoiding regions with large function values. For high-dimensional problems, as they appear for example in the context of topology optimization, a global optimization is often impossible due to the exponentially growing number of possible parameter values to test. In this case, one often resorts to a local optimization method that explores the parameter space starting from a given initial parameter vector~\cite{piggott2017fabrication,christiansen2018topology,sell2017large} or one considers a discretization of the parameter space and applies evolutionary optimization techniques~\cite{doi:10.1021/acsphotonics.6b00955,wiecha2017evolutionary}. Alternatively, one can recast the optimization problem \added{by searching for an optimal material distribution using algorithms such as ADMM or successive convex approximations, while treating the Maxwell equations as a constraint}\removed{and solve for the Maxwell equations and the optimal material distribution simultaneously}~\cite{Lu:11,Lu:13,piggott2015inverse}.

In this work, we focus on medium-dimensional optimization problems that allow for a global optimization \added{of parametrized geometries} based on the solution of Maxwell's equations, but do not allow for a complete scan of the parameter space. We consider two shape-optimization problems, i.e. the optimization of an integrated single-photon source and the optimization of a dielectric metasurface. Further, we consider the problem of a geometrical parameter reconstruction based on X-ray diffraction measurements.  

We benchmark optimization algorithms that are regularly applied in nano optics and that can be broadly assigned to three categories: local optimization, global stochastic optimization, and global model-based optimization.
\begin{itemize}
\item  Starting from a given initial parameter vector, {\bf local optimization methods} try to find better positions in the parameter space by exploring the neighborhood of the current position. They converge efficiently into a local minimum, which is not necessarily the global minimum. Gradient-based methods use first derivatives (gradients) or second derivatives (Hessians) in order to find a minimum in a smaller number of iterations. An example for a gradient-free method is the downhill simplex algorithm~\cite{doi:10.1093/comjnl/7.4.308} and an example for a gradient-based method is the low-memory Broyden-Fletcher-Goldfarb-Shanno (L-BFGS-B) algorithm~\cite{byrd1995limited}. 
The gradient of the solution to Maxwell's equations can be obtained by the \removed{adjoined}\added{adjoint} method~\cite{Lalau-Keraly:13,sell2017large} or by automatic differentiation~\cite{doi:10.1117/12.2270596}. Local optimization methods have been used, e.g., to optimize a Y-junction splitter~\cite{Lalau-Keraly:13} or a photonic nano-antenna~\cite{nikolay2017compact}, and to reconstruct geometrical parameters of a line grating from scatterometry data~\cite{doi:10.1117/12.2270596}.

\item {\bf Stochastic optimization algorithms} are based on random variables. Important representatives are particle swarm optimization~\cite{zhang2015comprehensive} and differential evolution~\cite{das2011differential}. These algorithms usually scale well for an increasing number of dimensions. However, they tend to require many function evaluations in order to converge to the global minimum. 
Particle-swarm optimization has been employed for optimizing diffraction grating filters~\cite{Shokooh-Saremi:07}, photonic-crystal waveguides~\cite{MIRJALILI20135989}, or the duality symmetry of core-shell particles~\cite{rahimzadegan2018core}. Differential evolution strategies have been investigated in the context of light focusing photonic crystals~\cite{bor2016differential} and for parameter extraction of optical materials~\cite{saber2017performance}. 

\item {\bf Model-based optimization methods} construct a model of the objective function in order to find promising sampling parameters. One important representative is Bayesian optimization, which constructs a statistical model of the objective function~\cite{shahriari2016taking}. 
Bayesian optimization is regularly used in machine learning applications~\cite{shahriari2016taking,Golovin:2017:GVS:3097983.3098043,aws.bayes}. In the field of nano-optics it has been employed to optimize ring resonator-based optical filters~\cite{Rehman:16}, chiral scatterers~\cite{Gutsche18}, and mantle cloaks~\cite{qin2018mantle}.
The method derives promising parameter samples by means of Bayesian inference based on \emph{all} previous function evaluations. This is in contrast to the other approaches, which only use few historic data points to determine new samples. While this statistical inference can drastically reduce the number of iterations, it requires a significant computational overhead on its own, which can slow down the optimization. We introduce an approach that aims to eliminate this slow-down for typical nano-optical computation scenarios. 

\end{itemize}

Another machine-learning technique that has been recently applied for nano-optical optimization is deep learning~\cite{malkiel2018plasmonic}. Trained with thousands of simulation results deep neural networks can serve as accurate models for mapping a geometry to an optical response and vice versa almost instantaneously. However, for this benchmark we only consider methods that do not require a training phase prior to the actual optimization.

\removed{The paper is organized as follows: In section~\ref{sec:optimization_methods} the considered optimization methods and their implementation are introduced and strategies for their parallelization are described. The three optimization problems are introduced in section~\ref{sec:optimization_problems}. After the presentation and discussion of the numerical experiments in section \ref{sec:numerical_experiments} the paper concludes in section \ref{sec:conclusion}.}

\section*{Examined optimization methods}
\label{sec:optimization_methods}

In the field of optical simulations one has often access to computing clusters or powerful multi-core computers. It is therefore important that optimization methods exploit the possibility of computing the objective function for several input values in parallel. Moreover, it should be possible to distribute the computation of the objective function to several machines or threads. To this end, we have integrated the considered optimization methods in a server-client framework following the design strategy of Google Vizier~\cite{Golovin:2017:GVS:3097983.3098043}. Furthermore, we adapted the optimization methods to support inequality constraints of the parameter space that arise from geometrical or practical (e.g. fabrication) requirements.

In the following, we shortly describe the optimization approaches. Further details on the numerical framework, the implementation of the algorithms supporting a constrained optimization as well as a visualization of the different optimization strategies are contained in the Supporting information. 

\subsection*{Local optimization methods}

In local optimization methods, the next sampling point depends on the function value of the previous sample. A parallelization of local optimization methods is achieved by starting several independent local optimizations from different pseudo-random points in the parameter space $\mathcal{X} \subset \mathbb{R}^D$. \added{After a local optimization has converged, it is restarted from a different point such that eventually the global minimum would be found.}

For optimization problems that do not exploit derivative information, we consider the downhill simplex algorithm~\cite{doi:10.1093/comjnl/7.4.308}. Otherwise, we consider the gradient-based L-BFGS-B algorithm~\cite{byrd1995limited}. Both methods are implemented based on the python package \texttt{scipy.optimize}~\cite{scipy}.

\subsection*{Stochastic global optimization}

As stochastic global optimization methods, we consider particle swarm optimization and differential evolution.

Particle swarm optimization works by randomly moving the position of each particle in the search-space guided by the particle's best known position as well as the swarm's best known position. The method is implemented based on the Python package \texttt{pyswarm}~\cite{pyswarm}, which supports a parallel evaluation of the objective function. 

Differential evolution is a population-based genetic algorithm that is implemented based on the python package \texttt{scipy.optimize.differential\_evolution}~
\cite{scipy,scipy.differentialevolution}. We extend the algorithm by allowing for a parallel evaluation of the fitness function for each offspring.  

The performance of the stochastic global optimization algorithms heavily depends on parameters such as the swarm size and the population size (see supporting information for a comparison). Due to the computational complexity of adapting these parameters, the benchmark is performed with the default parameters of the python packages.

\subsection*{Bayesian optimization}

Bayesian optimization is based on a stochastic model of the objective function. We employ a Gaussian process (GP) that is updated with each new evaluation of the objective function through GP regression (a form of Bayesian inference)~\cite{williams1998prediction,shahriari2016taking}. The GP allows to identify parameter values with the largest expected improvement ${\rm EI}(\mathbf{x})$. By sampling those points, exploitation and exploration are automatically balanced, i.e. after exploiting a possible improvement within a local minimum, other parts of the parameter space are automatically scanned until a better minimum is found. As proposed by Gonzales {\em et al.}~\cite{pmlr-v51-gonzalez16a} the parallelization is realized by penalizing the expected improvement of the parameters close to running calculations.

 One important advantage of GP regression is that it does not rely on derivative information on the objective function, but that one can incorporate this information if available~\cite{solak2003derivative}. We use an in-house implementation of Bayesian optimization that exploits the possibility to include derivative observations~\cite{hammerschmidt2018solving}. As will be shown below, this can speed up the optimization significantly.

The search process for the next sampling point is itself an optimization problem that can be computationally demanding and may slow down the optimization. To tackle this problem in the context of parallel computations of the objective function, we use two strategies: (i) The next sampling point is computed in advance, i.e. in parallel to the evaluation of the objective functions. (ii) We use differential evolution to maximize the expected improvement and automatically adapt the effort to the average time interval at which new objective function values are acquired. See the supporting information for a detailed description of the approach.

\section*{Optimization Problems}
\label{sec:optimization_problems}

For benchmarking the different optimization algorithms, we consider three technologically relevant optimization problems of contemporary interest: the maximization of the coupling efficiency of a single-photon source to an optical fiber, the parameter reconstruction of a lamellar grating from scattering data, and the reflection suppression from a silicon metasurface. These three problems are described in the following individually prior discussing the results from the optimization.

\subsection*{Improving the coupling efficiency of a single-photon source}

\begin{figure*}[ht]
\center
\def\svgwidth{0.3\textwidth}
\raisebox{4.0cm}{a)}\tiny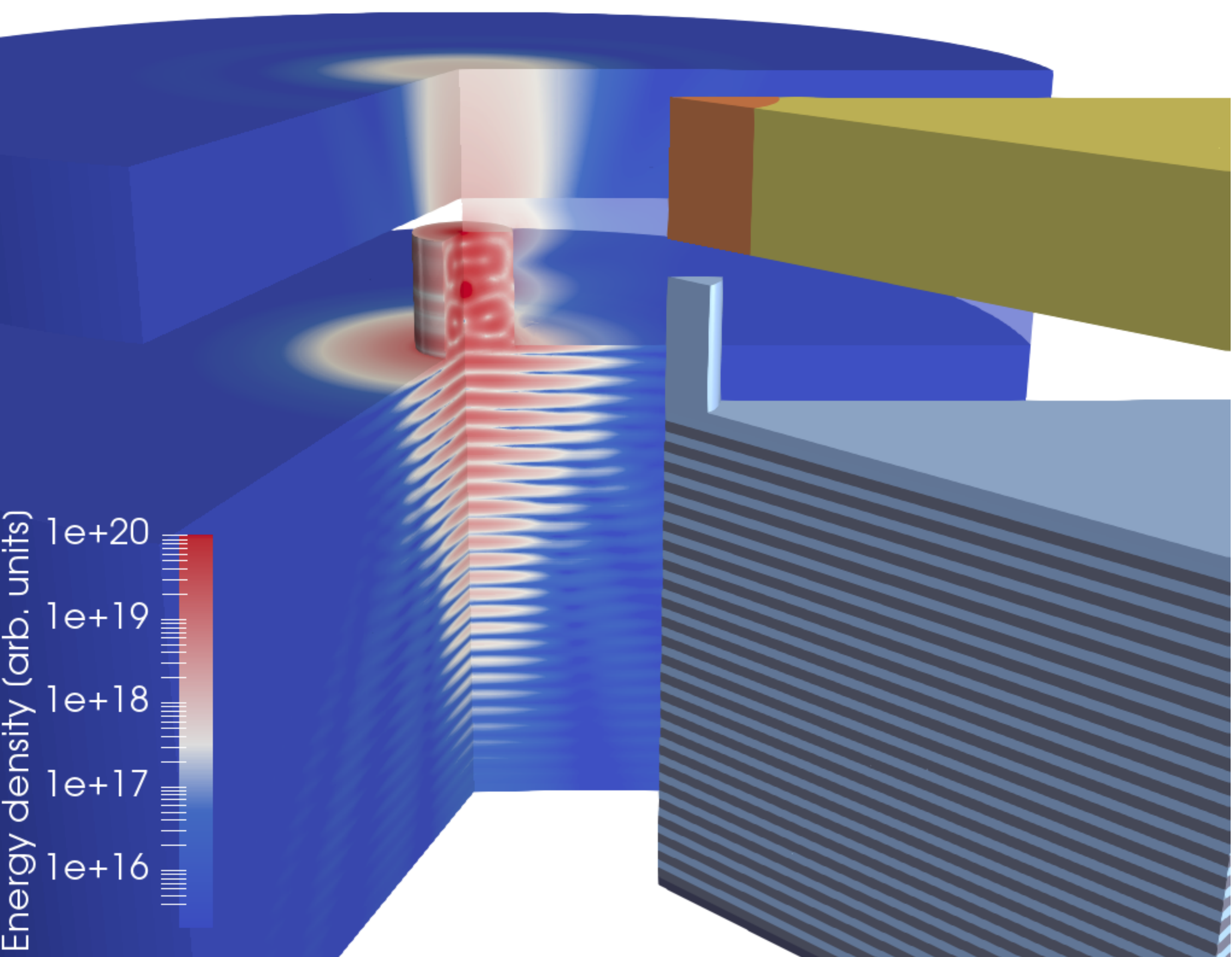\normalsize
\raisebox{4.0cm}{b)}\includegraphics[width=0.31\textwidth]{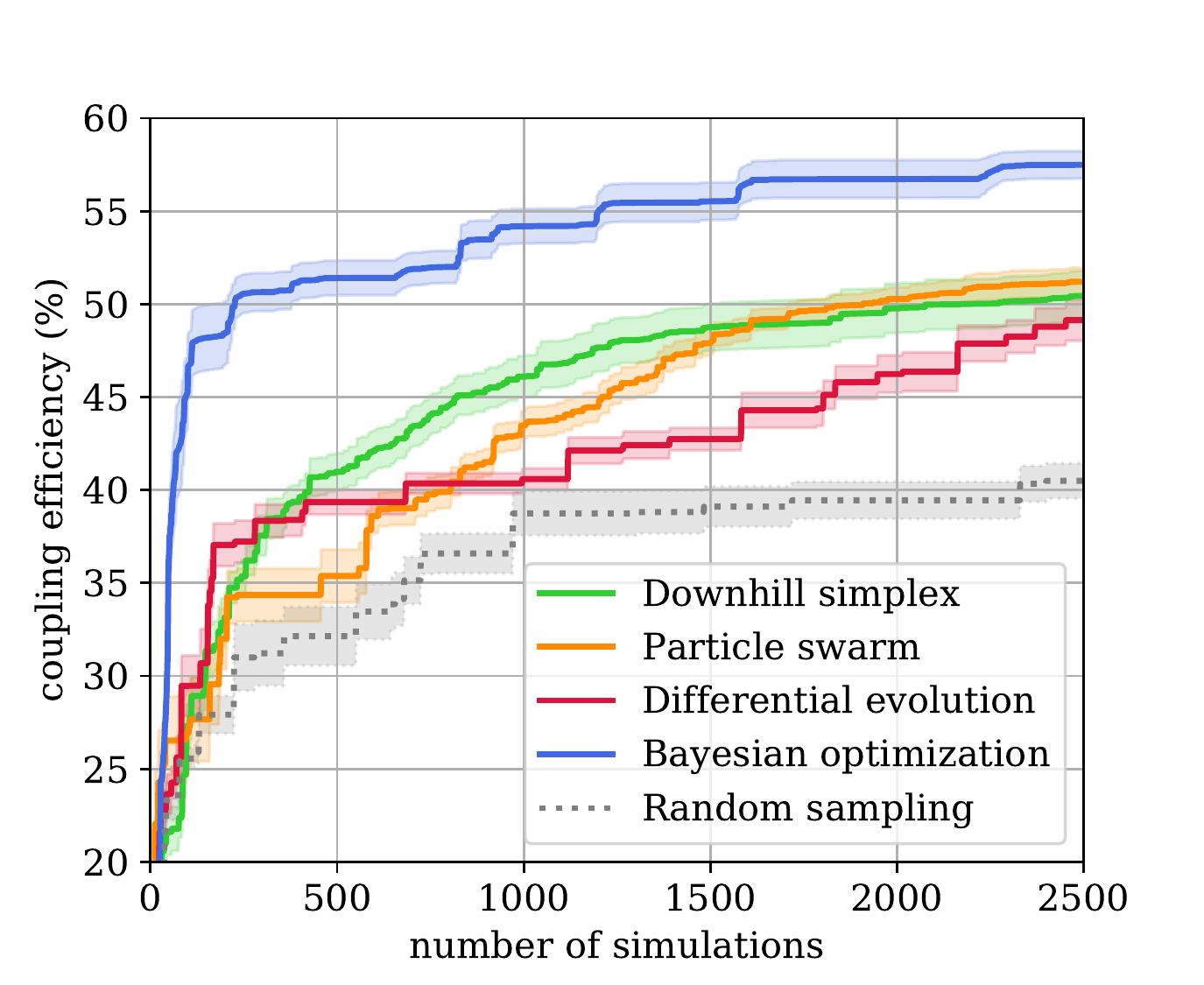}
\raisebox{4.0cm}{c)}\includegraphics[width=0.31\textwidth]{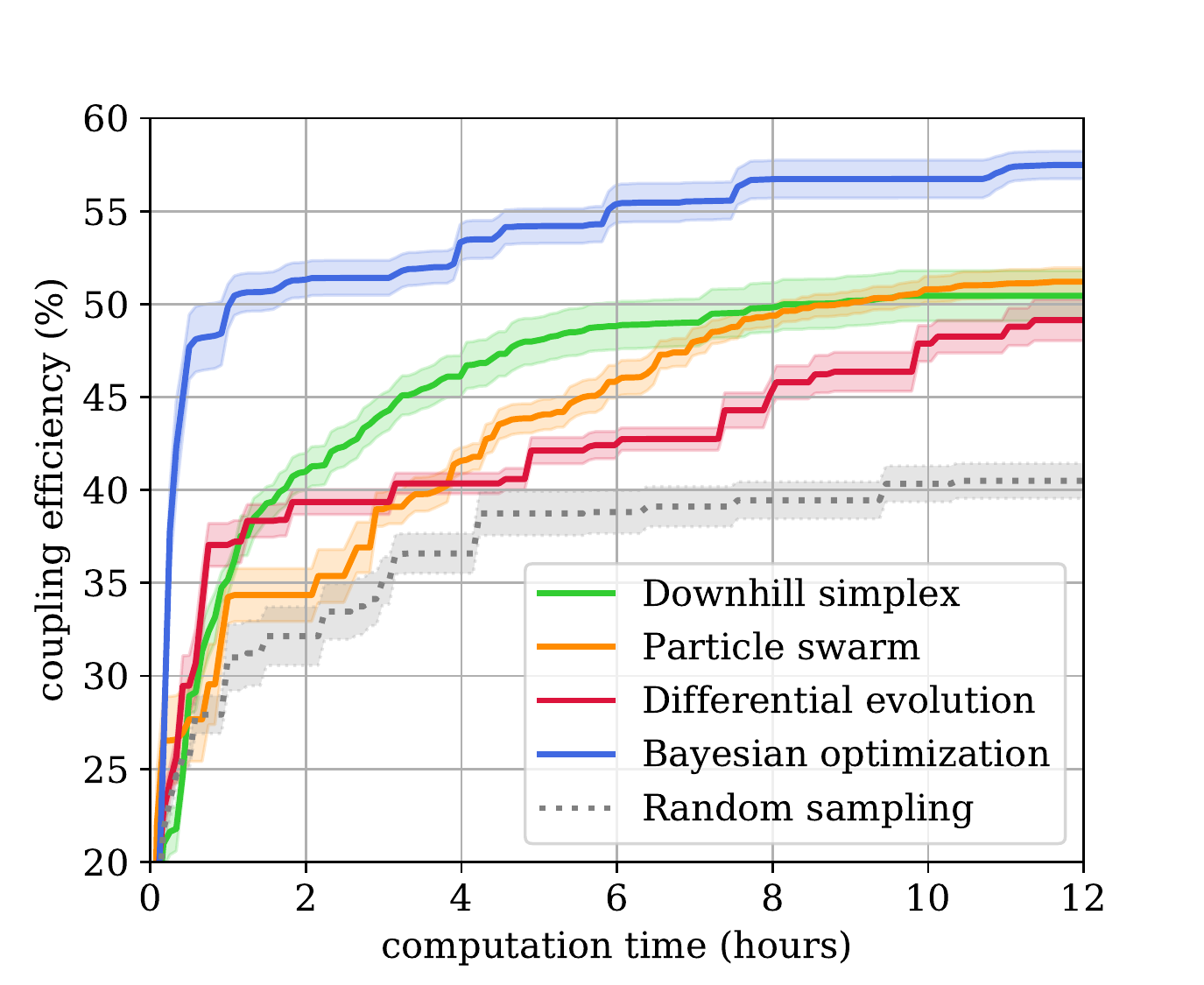}
\caption{
{\bf a)} Visualization of the energy density of the light field of the optimized fiber-coupled single-photon source with a coupling efficiency of 60\%. A cut through the geometry is shown in front of the energy-density plot. 
The single-photon source consists of a QD dipole source embedded into a mesa structure (GaAs, blue), a Bragg reflector [alternating layers of GaAs (blue) and \AlGaAs~(gray)], and an optical fiber with homogeneous fiber core (orange) and fiber cladding (yellow). The Bragg reflector is grown on a substrate made of GaAs and has a GaAs top layer (blue). The six optimized parameters are the mesa height $h_{\rm mesa} = 1217\,{\rm nm}$, 
mesa \added{radius $r_{\rm mesa} = 522\,{\rm nm}$}\removed{width $w_{\rm mesa} = 1044\,{\rm nm}$}, 
top-layer thickness $h_{\rm layer} = 190\,{\rm nm}$,
dipole elevation within the mesa $h_{\rm dip} = 613\,{\rm nm}$,
fiber-core \added{radius $r_{\rm core} = 969\,{\rm nm}$} \removed{diameter $d_{\rm core} = 1937\,{\rm nm}$}, and
mesa-fiber distance $s_{\rm mf} = 352\,{\rm nm}$.
{\bf b)}: Best seen coupling efficiency for different optimization approaches as a function of the number of simulations averaged over six independent optimization runs. The shading indicates the \added{standard error, i.~e. the} uncertainty of the average. 
{\bf c)}: Same as b) but the best seen coupling efficiency is shown as a function of the total computation time. The comparison with a) shows that Bayesian optimization has no significant computational overhead compared to the other optimization approaches.}
\label{fig:SPE}
\end{figure*}

Single-photon sources are essential building blocks of future photonic and quantum optical devices. We consider a source consisting of a quantum dot (QD) emitting at a vacuum wavelength of $\lambda=1,300\,$nm in the telecom O-band. The QD is embedded into a mesa structure made from gallium arsenide (GaAs; refractive index $n_{\rm GaAs} = 3.4$). An underlying Bragg reflector made from layers of GaAs and aluminum gallium arsenide (\AlGaAs; $n_{\rm AlGaAs}=3.0$) reflects the light emitted by the QD back into the upper hemisphere. The light is coupled into an optical fiber with large numerical aperture (NA) above the QD consisting of a homogeneous fiber core and a homogeneous fiber cladding ($n_{\rm core} = 1.5$, $n_{\rm clad} = 1.45$, ${\rm NA} \equiv \sqrt{n_{\rm core}^2-n_{\rm clad}^2} = 0.38$). The setup is sketched in Fig.~\ref{fig:SPE} a).

The parameter space $\mathcal{X}$ is spanned by 6 parameters: the height of the top layer above the Bragg reflector ($h_{\rm layer}$), the \added{radius}\removed{diameter} of the fiber core (\added{$r_{\rm core}$}\removed{$d_{\rm core}$}), the \added{radius}\removed{width} and height of the mesa (\added{$r_{\rm mesa}$}\removed{$w_{\rm mesa}$}, $h_{\rm mesa}$), the elevation of the dipole within the mesa ($h_{\rm dip}$), and the distance between mesa and fiber ($s_{\rm mf}$). The objective of the optimization is to maximize the coupling efficiency of the emitted light into the fundamental modes of the fiber. The numerical method to determine the coupling efficiency was described by Schneider {\em et al.}~\cite{Schneider:18}.

\subsection*{Parameter reconstruction of a lamellar grating}

\begin{figure*}[ht]
\centering
\begin{minipage}[b]{0.33\textwidth}
	
	\def\svgwidth{0.93\textwidth}
	\raisebox{2cm}{a)}\tiny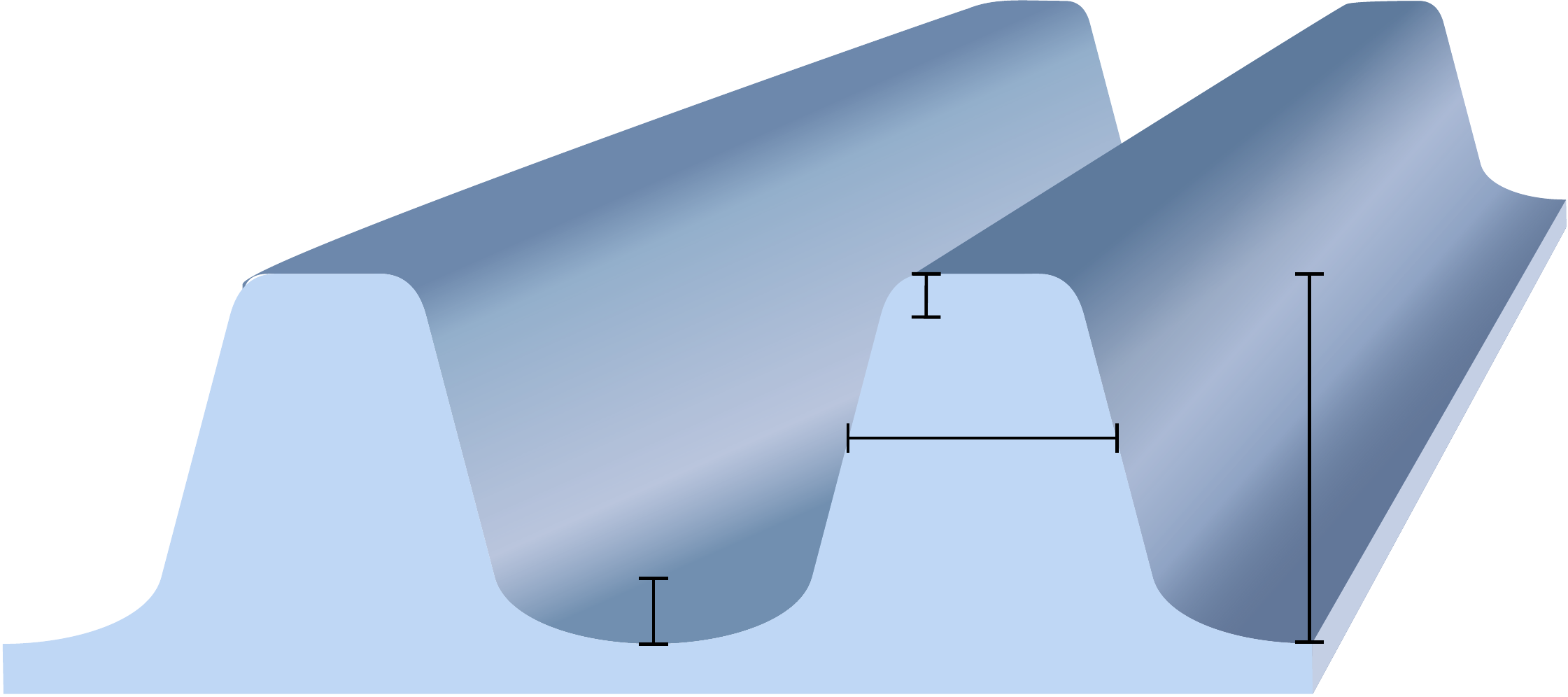\normalsize
	\def\svgwidth{0.85\textwidth}
	\raisebox{3.0cm}{b)}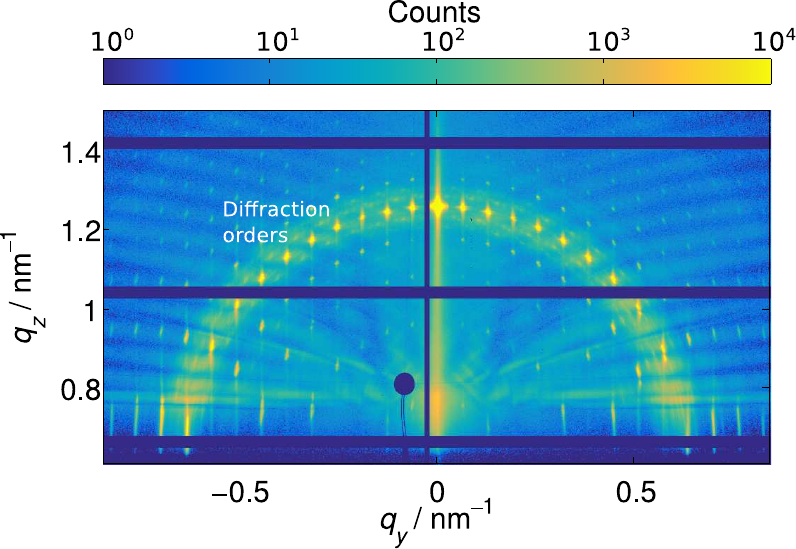
\end{minipage}
\hfill
\begin{minipage}[b]{0.32\textwidth}
\raisebox{0.4cm}{~c)}
\hspace*{-0.5cm}\includegraphics[width=1.0\textwidth]{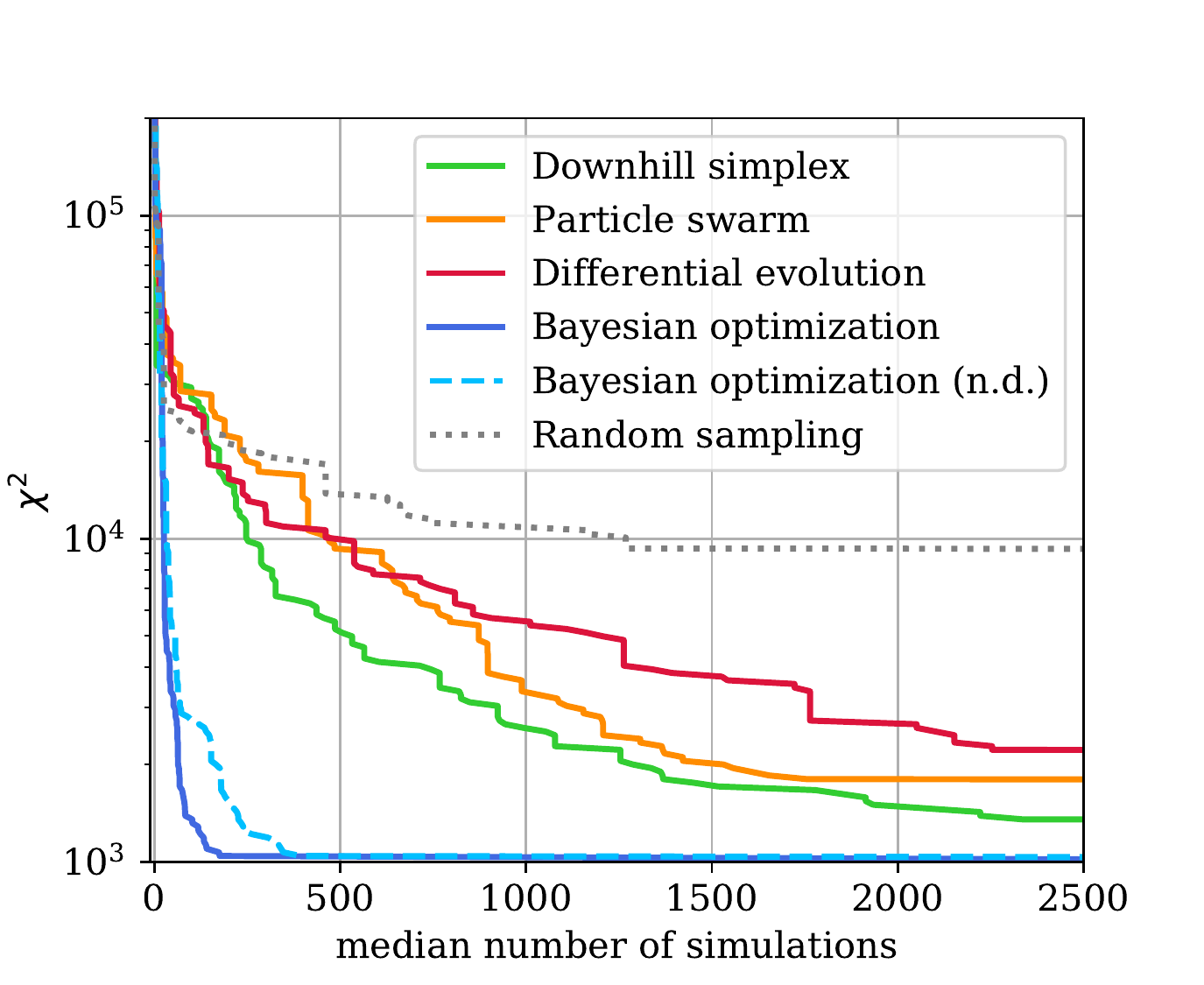}
\end{minipage}
\begin{minipage}[b]{0.32\textwidth}
\raisebox{0.4cm}{~d)}
\hspace*{-0.5cm}\includegraphics[width=1.0\textwidth]{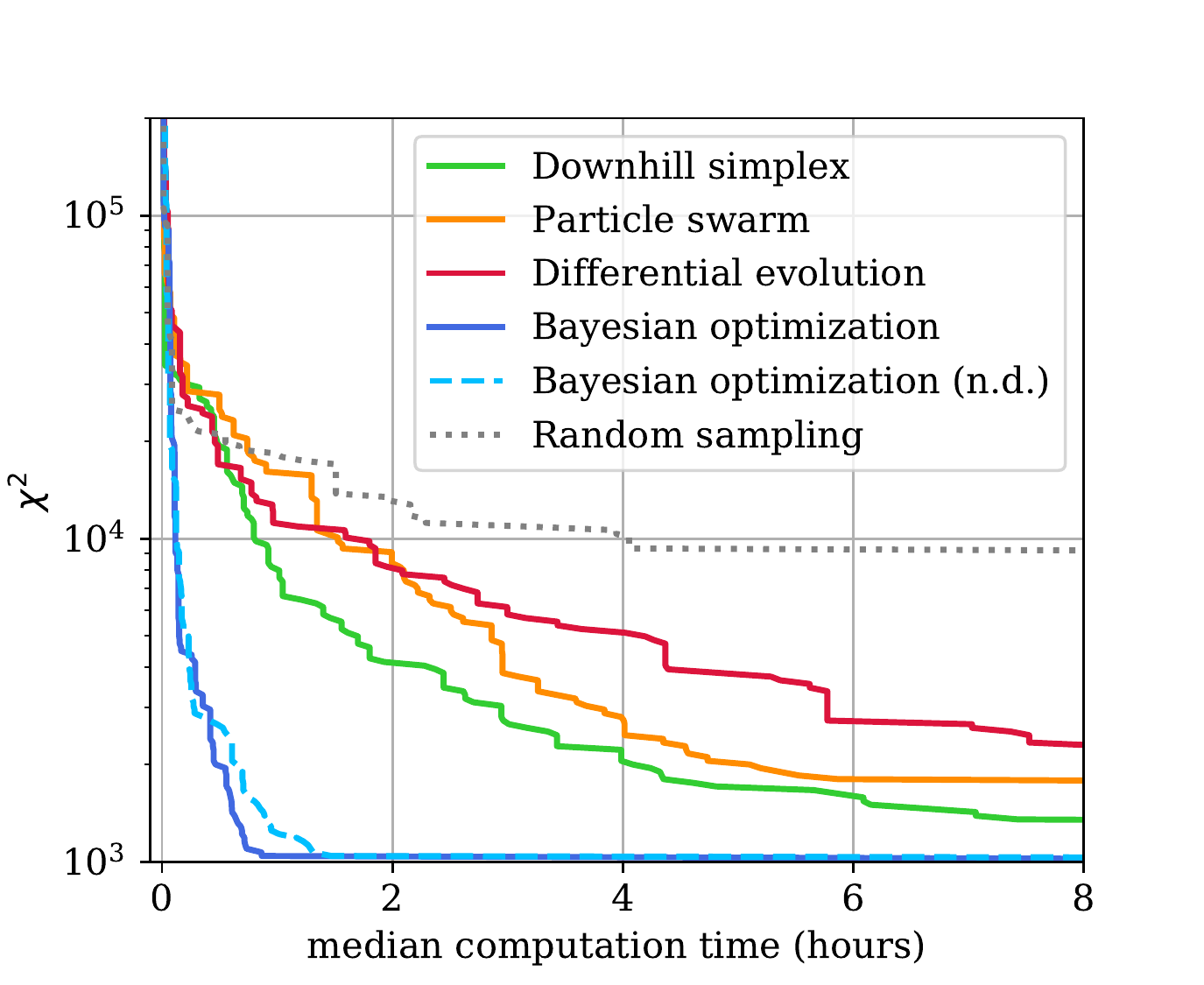}
\end{minipage}
\caption{
{\bf a)} Geometry of the lamellar grating. 
The aim of the GISAXS measurement is to reconstruct the critical dimension ${\rm CD}$, the line height $h_{\rm line}$, the side-wall angle $\phi_{\rm swa}$, the top corner radius $r_{\rm top}$, and the depth of the grooves $d_{\rm groove}$. 
{\bf b)} GISAXS scattering pattern of the grating obtained at a photon energy of 6.5\,keV shown as a function of the scattering vector components $\mathbf{q}_y$ and $\mathbf{q}_z$. The intensity is shown with a logarithmic false color scale.
{\bf c)}: Error $\chi^2$ [see Eq.~\eqref{eq:chi_squared}] for different optimization approaches as a function of the median number of simulations needed to reach a value below $\chi^2$ for six independent optimization runs. The blue and green lines show the results of Bayesian optimization including derivatives and using no derivative (n.d.) information, respectively. 
{\bf d)}: Same as c) but $\chi^2$ is shown as a function of the median total computation time.
}
\label{fig:Gisaxs}
\end{figure*}

Grazing incidence small angle X-ray scattering (GISAXS) is a destruction-free scatterometry method. With incidence angles close to the critical angle of total external reflection, GISAXS is a technique with high surface sensitivity. We consider the parameter reconstruction of a periodic, lamellar silicon grating manufactured using electron beam lithography~\cite{soltwisch2017reconstructing}. The grating geometry is modeled by 5 parameters: the critical dimension ${\rm CD}$ (line width), the line height $h_{\rm line}$, the side-wall angle $\phi_{\rm swa}$, the top corner radius $r_{\rm top}$, and the depth of the grooves $d_{\rm groove}$ [see Fig.~\ref{fig:Gisaxs} a)]. 
A monochromatic X-ray beam idealized as a plane wave
with the wave vector $\mathbf{k}_{\rm in}$ impinges on the sample surface at a grazing incidence angle. The elastically scattered wave with the wave vector $\mathbf{k}_{\rm f}$ propagates to a 2D area detector that records a scattering intensity pattern $I_{\rm det}(\mathbf{q})$ as a function of the scattering vector $\mathbf{q} = \mathbf{k}_{\rm f} - \mathbf{k}_{\rm in}$ [see Fig.~\ref{fig:Gisaxs} b)]. Based on a FEM model of the experiment, one can determine a scattering intensity $I_{\rm FEM}(\mathbf{p},E,N)$ for each parameter set $\mathbf{p}\in\mathcal{X}$, photon energy $E$, and diffraction order $N$~\cite{soltwisch2017reconstructing}.

The natural line edge and line width roughness of the grating is taken into account with an analytic approach based on Debye-Waller damping
$\exp(-[\sigma_{\rm r} \mathbf{q}_y(N)]^2)$ with the damping factor $\sigma_{\rm r}$ and the $y$-component $\mathbf{q}_y(N)$ of the scattering vector for diffraction order $N$. Furthermore, the model intensities are scaled by an energy-dependent factor $s(E)$ in order to account for the unknown effective illumination area of the sample. This leads to the model $I_{\rm model}(\mathbf{p},E,N) = I_{\rm FEM}(\mathbf{p},E,N)\, s(E)\, \exp(-[\sigma_{\rm r} \mathbf{q}_y(N)]^2)$
for the measured intensities in each diffraction order $N$.

The data acquisition is performed for three different photon energies, $E_1 = 5.5\,{\rm kEV}$, $E_2 = 5.55\,{\rm kEV}$, and $E_3 = 5.6\,{\rm kEV}$. For each energy, the standard deviation $\sigma(E_i)$ of the measured intensities was determined, as described by Soltwisch {\em et al.}~\cite{soltwisch2017reconstructing}. The aim of the optimization is to minimize the mean squared numerical error of all diffraction orders and energies
\begin{equation}
\label{eq:chi_squared}
\chi^2 = \sum_{i=1}^3 \sum_{N=1}^M \frac{\left(
I_{\rm model}(\mathbf{p},E_i,N)
- I_{\rm exp}(E_i,N)
\right)^2}{\sigma^2(E_i)}
\end{equation}
with respect to the geometrical parameters ${\rm CD}$, $h_{\rm line}$, $\phi_{\rm swa}$, $r_{\rm top}$, $d_{\rm groove}$, the roughness $\sigma_{\rm r}$, and the three scaling factors $s_i = s(E_i)$ for $i=1,2,3$.

We note, that it is straightforward to determine the derivatives of $\chi^2$ with respect to the parameters $\sigma_{\rm r}$, $s_1$, $s_2$, and $s_3$ that do not enter into the numerical FEM simulation of the measurement process. In the next section we will make use of these derivative information in order to assess to which extend partial information on parameter derivatives can speed up the Bayesian optimization process.

\subsection*{Reflection suppression by a metasurface}

\begin{figure*}[ht]
\centering
\raisebox{3.8cm}{a)}\def\svgwidth{0.33\textwidth}
\tiny
\hspace*{-0.5cm}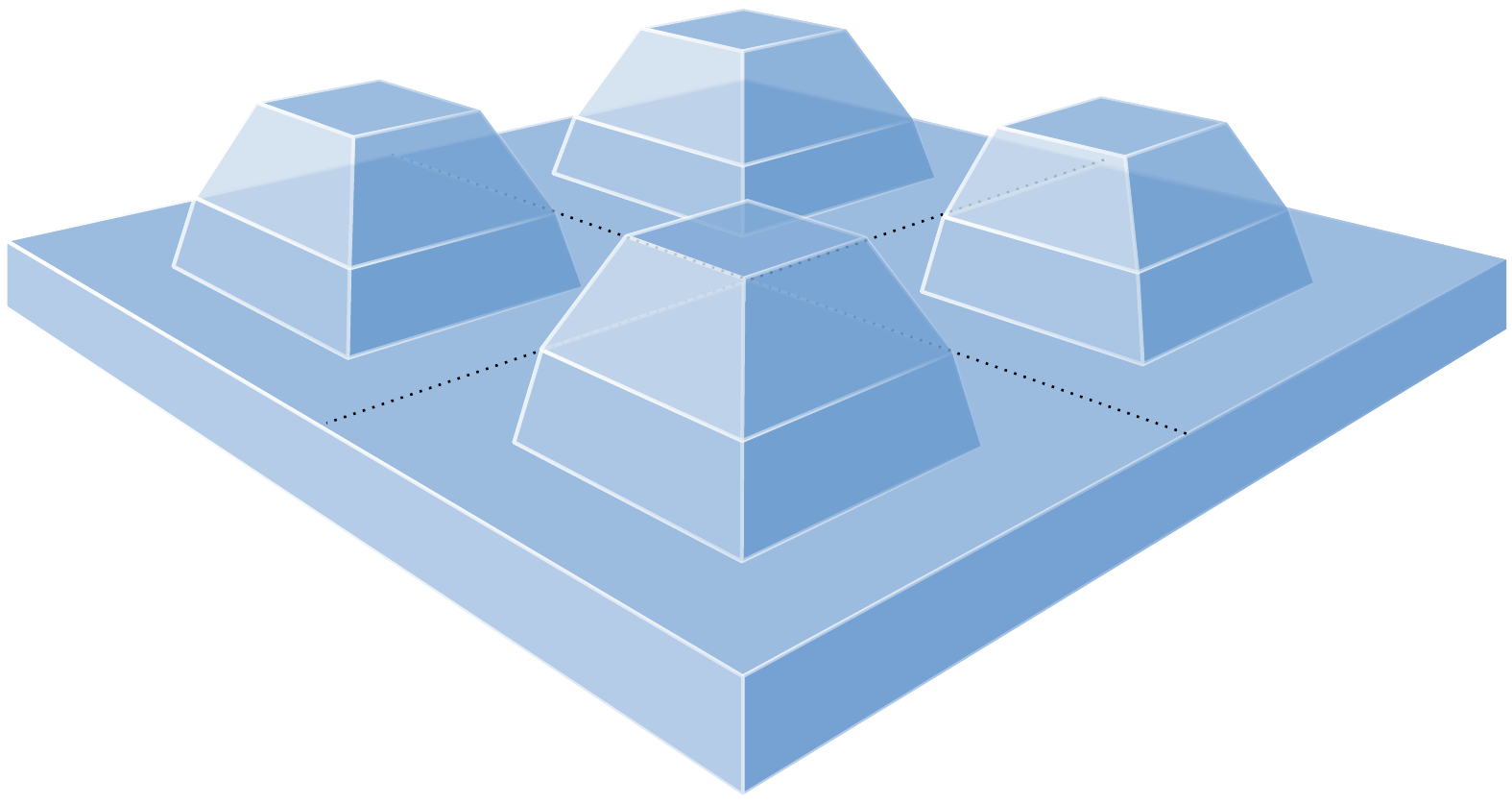
\normalsize
\hspace*{-0.2cm}\raisebox{3.8cm}{b)}
\hspace*{-0.17cm}\includegraphics[width=0.31\linewidth]{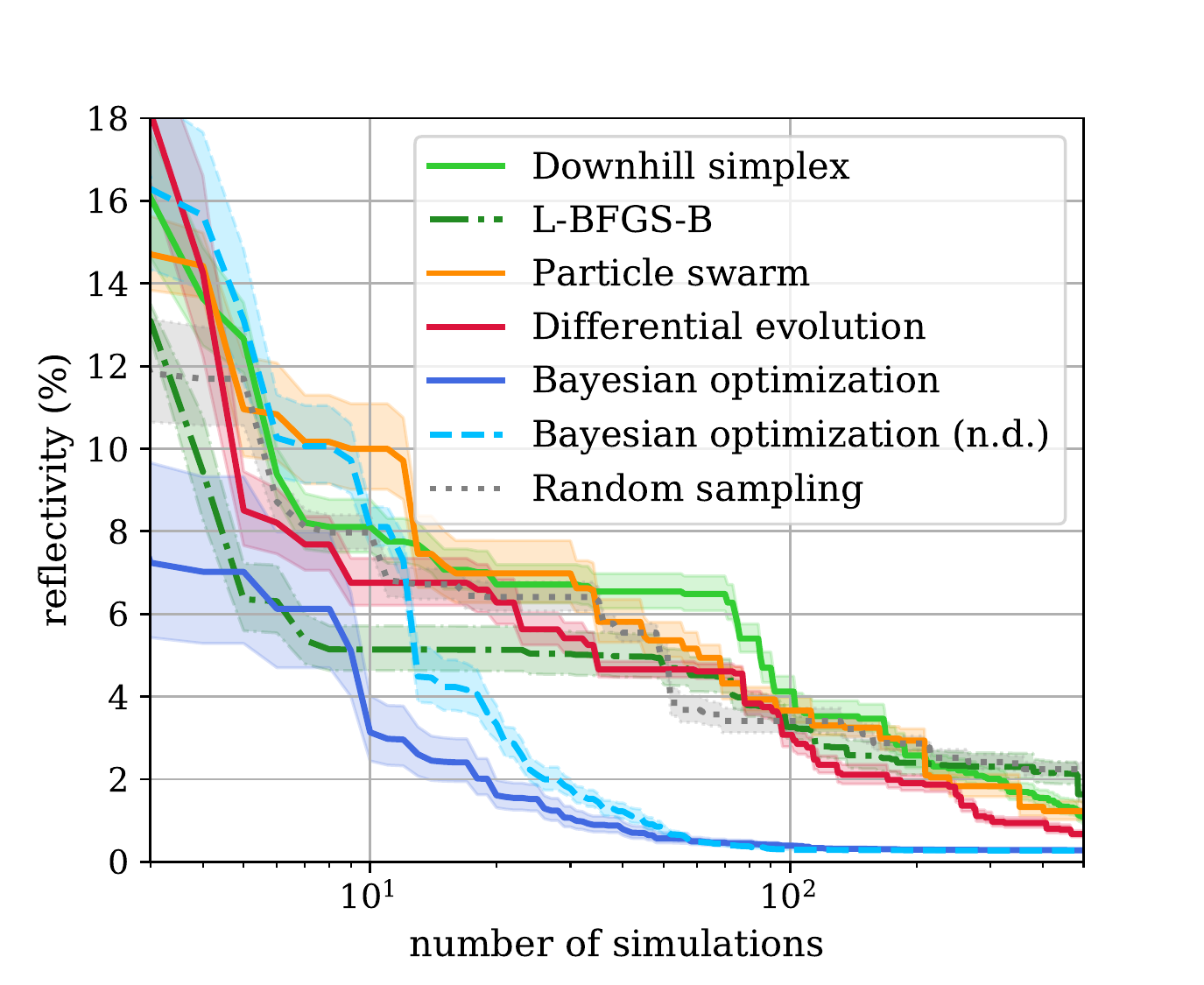}
\hspace*{-0.2cm}\raisebox{3.8cm}{c)}
\hspace*{-0.17cm}\includegraphics[width=0.31\linewidth]{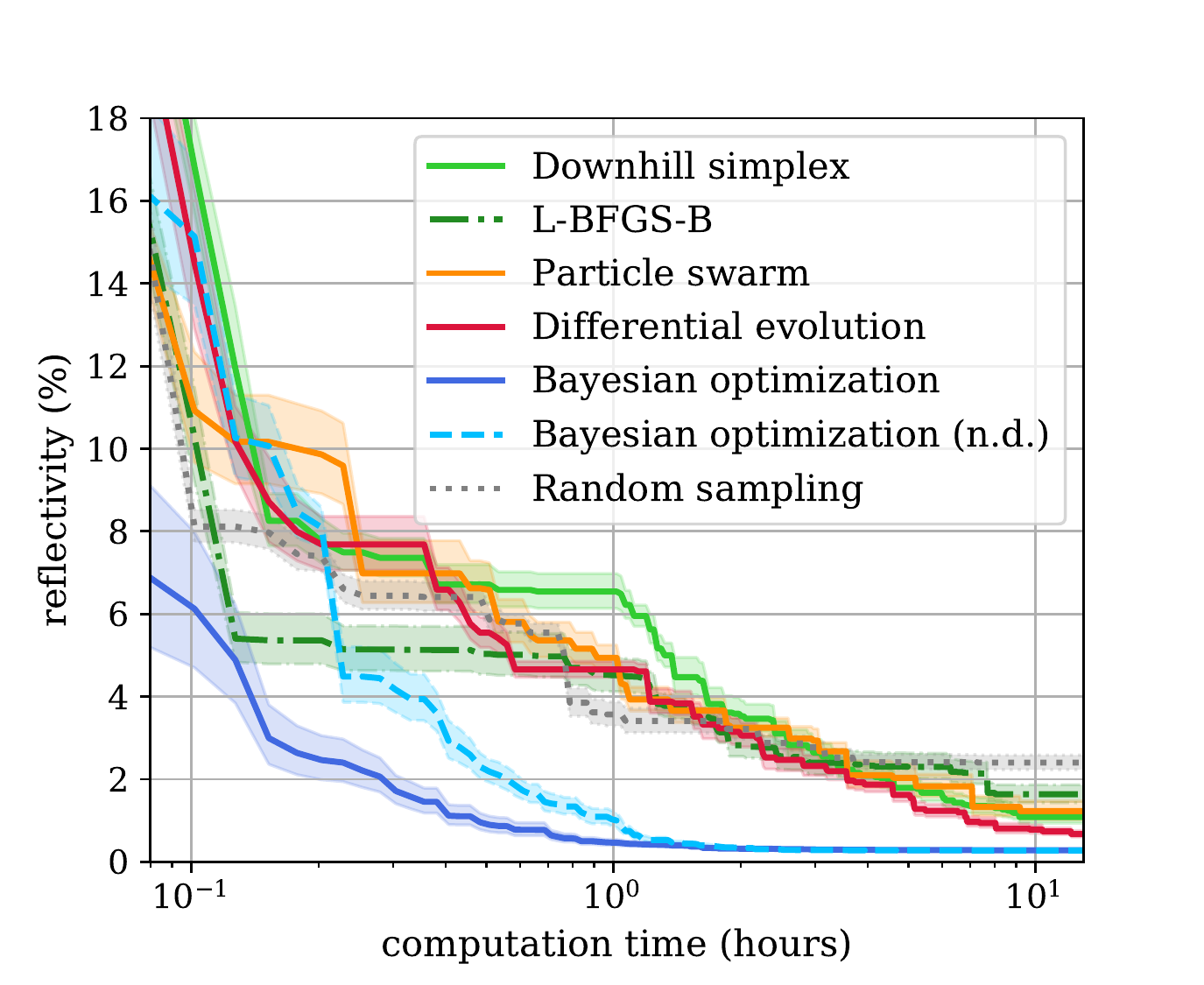}
\caption{
{\bf a)}: The metasurface consists of a square array of silicon bumps on a silicon substrate. The reflectivity of the structure is minimized with respect to the periodicity of the array ($p$), 
the heights of the lower and upper part of the bumps ($h_{\rm lower}$ and $h_{\rm upper}$), and the bottom, middle, and top width of the bumps ($w_{\rm bottom}$, $w_{\rm middle}$, and $w_{\rm top}$).
{\bf b)}: Lowest reflectivity for different optimization approaches as a function of the number of simulations averaged over six independent optimization runs. The shading indicates the \added{standard error.}\removed{uncertainty of the average.} 
{\bf c)}: Same as b) but the lowest reflectivity is shown as a function of the total computation time.
Bayesian optimization using no derivative information ["Bayesian optimization (n.d.)"] takes roughly twice as long to obtain a reflectivity below 1\%.
}
\label{fig:Meta}
\end{figure*}

Broadband anti-reflection is a desirable property, e.g., for high-efficiency solar cells as well as for CCD or CMOS sensors. We consider a nano-structured silicon metasurface for suppressing the reflectivity in a broad range of frequencies~\cite{proust2016optimized}. The metasurface consists of a square array of silicon bumps on top of a silicon substrate [see Fig.~\ref{fig:Meta} a)] and is parametrized by six length scales: the periodicity of the array ($p$), the heights of the lower and upper part of the bumps ($h_{\rm lower}$, $h_{\rm upper}$), and the bottom, middle, and top width of the bumps ($w_{\rm bottom}$, $w_{\rm middle}$, $w_{\rm top}$). The reflectivity of incident plane waves perpendicular to the surface is averaged over four wavelengths, $500\,{\rm nm}$, $600\,{\rm nm}$, $700\,{\rm nm}$, and $800\,{\rm nm}$. In order to facilitate the fabrication of the structure we constrain the parameter space of the optimization such that $w_{\rm bottom} \leq p - 10\,{\rm nm}$, $w_{\rm bottom} \leq w_{\rm middle} \leq w_{\rm top}$ and $h_{\rm lower} + h_{\rm upper} \leq 2 w_{\rm bottom}$. 

Using automatic differentiation, the FEM software JCMsuite~\cite{JCMsuite} allows to determine partial derivatives of the computed fields with respect to geometrical parameters with small computational overhead. Based on this information, we compute for this example the partial derivatives of the reflectivity with respect to the six design parameters.  
 
\section*{Numerical experiments}
\label{sec:numerical_experiments}

The optimization runs for the three optimization problems introduced in the previous section have been each performed for a fixed number of iterations (single-photon source optimization and parameter reconstruction 2,500 iterations, anti-reflective metasurface optimization 500 iterations). In order to evaluate the average performance of the different optimization methods, each run has been repeated 6 times with different initial conditions. All methods were run with typical numerical settings (see supporting information). A targeted adaptation of the settings to a specific optimization problem is generally computationally much more demanding than the optimization problem itself~\cite{pedersen2010tuning}. In practice, the problem-adapted performance is therefore usually less relevant than the performance with default parameters.

In addition to the optimization methods introduced in the first section, results for a random sampling of the objective functions are also considered. If an optimization method is not significantly better than the non-informative random sampling, it suffers from the specific properties of the objective function, e.g. irregularities or the existence of too many local minima.

The optimization of the single-photon source and the parameter reconstruction were performed on a machine with a 6-core Intel Xeon CPUs running at 3.2 GHz with 11 GB of RAM. The optimization methods were configured to perform at most 4 parallel computations of the objective function. The optimization of the metasurface is numerically much more demanding as a three-dimensional geometry is considered and a wavelength scan has to be performed. Therefore, the optimizations were run on a more powerful machine with 4 Intel Xeon CPUs with 10 cores at 2.4 GHz with 1 TB of RAM. Up to 6 parallel computations of the objective function were performed. On both machines Bayesian optimization used four parallel threads for computing the sampling points.

\subsection*{Optimized single-photon source}
\label{ssec:NumSPE}

Figure~\ref{fig:SPE} b) and c) compares the performance of Bayesian optimization, particle swarm optimization, differential evolution, the downhill simplex method, and random sampling for the example of the maximization of the coupling efficiency of the emitted light into the optical fiber. Bayesian optimization clearly outperforms all other methods. After about 1,200 iterations and a computation time of 6 hours it reaches an average coupling efficiency of 55\% while the other optimization methods reach only average efficiencies below 52\% after 2,500 iteration and a computation time of 12 hours. A comparison of Figs.~\ref{fig:SPE} b) and c) shows that Bayesian optimization has no significant computational overhead compared to the other optimization approaches.

The computed optimal configuration with
mesa height $h_{\rm mesa} = 1217\,{\rm nm}$, 
mesa \added{radius}\removed{width} $r_{\rm mesa} = 522\,{\rm nm}$, 
top-layer thickness $h_{\rm layer} = 190\,{\rm nm}$,
dipole elevation $h_{\rm dip} = 613\,{\rm nm}$,
fiber-core \added{radius}\removed{diameter} $r_{\rm core} = 979\,{\rm nm}$, and
mesa-fiber distance $s_{\rm mf} = 352\,{\rm nm}$ is shown in Fig.~\ref{fig:SPE} a). The setup achieves a remarkably large coupling efficiency of 60\% to the optical fiber. In a previous work, we tried to improve the coupling efficiency of the same system by performing numerical scans of all parameters of the geometry apart from the top-layer thickness $h_{\rm layer}$, which was fixed to 195\,nm. A full parameter scan of all five parameters with only 10 values per parameter would have required 100,000 simulations. Therefore, we restricted in the prior work the scan to a small fraction of the physically realizable parameter space. For example, dipole elevations only between 0 and 50\,nm in steps of 10\,nm were considered, while the current optimization considers values between 0 and 1000\,nm. Due to this restriction, a maximal coupling efficiency of only 23\% had been obtained in the past~\cite{Schneider:18}. This demonstrates that an optimization within a large space of realizable system parameters is important to assess the technological potential of a nano-optical system.

\subsection*{Parameter reconstruction of a lamellar grating}

For practical applications of shape optimization it is of interest to determine a reasonably good structure in a limited time budget (e.g. in one day). In contrast, in the context of a parameter reconstruction it is important to find parameter values close to the \emph{global} optimum with a small error level $\chi^2$ as fast as possible. Correspondingly, Fig.~\ref{fig:Gisaxs} c) and d) show the \emph{median} number of simulations and the \emph{median} time span needed to reach a specific error level $\chi^2$. 
The optimal parameters correspond to the values obtained by Soltwisch {\em et al.}~\cite{soltwisch2017reconstructing}.

In this case, Bayesian optimization shows an impressive lead in comparison with the other optimization methods. Using the derivative information with respect to the parameters $\sigma_{\rm r},s_1,s_2,s_3$, Bayesian optimization reaches an error level of $10^3$ after a median number of about 200 simulations and a median time of one hour (blue line), while the non-Bayesian methods do not manage to reach the same error level even after 2,500 iterations or eight hours computation time.

\subsection*{Optimized anti-reflective metasurface}

For the example of the reflection suppression using a metasurface, all first order partial derivatives with respect to the six geometrical parameters were determined by automatic differentiation. This enables to apply also the gradient-based L-BFGS-B method. The benchmark results are presented in Figs.~\ref{fig:Meta} b) and c). Again Bayesian optimization, even without using derivative information ["Bayesian optimization (n.d.)"], optimizes the structure significantly faster. It reaches average reflectivities below 0.5\% while Differential evolution obtaines 0.7\% and the other optimization methods between 1\% and 2\%. The use of derivatives for Bayesian optimization offers again a significant advantage. It is remarkable that this is not the case for L-BFGS-B optimization. After a relatively fast initial optimization to 5\% reflectivity, L-BFGS-B performs comparable to the other non-derivative methods, including random sampling. We attribute this to a complex shape of the objective function with many similarly low local minima.

\begin{figure}[htb]
\raisebox{4.2cm}{a)}
\hspace*{-0.1cm}\includegraphics[width=0.28\linewidth]{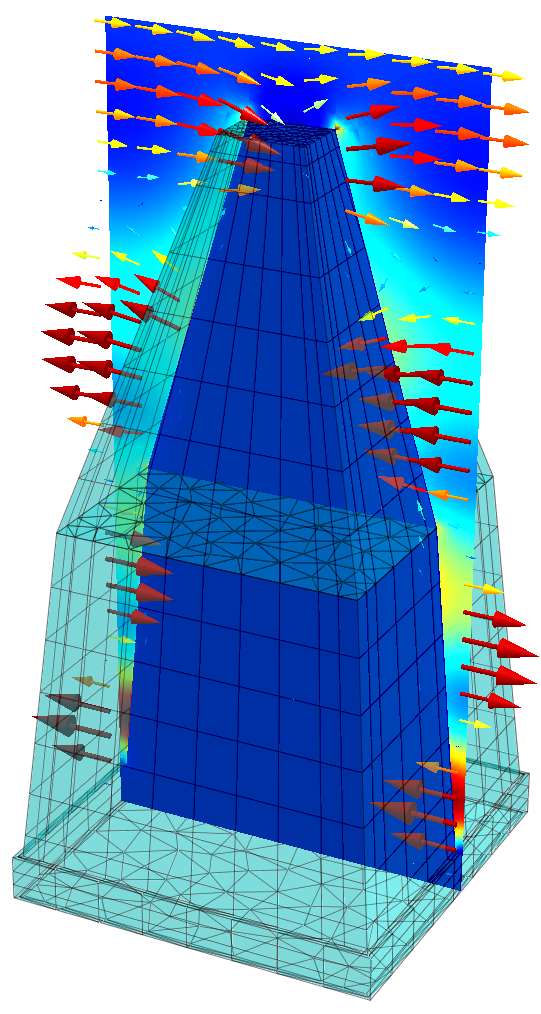}
\raisebox{4.2cm}{b)}
\hspace*{-0.1cm}\includegraphics[width=0.58\linewidth]{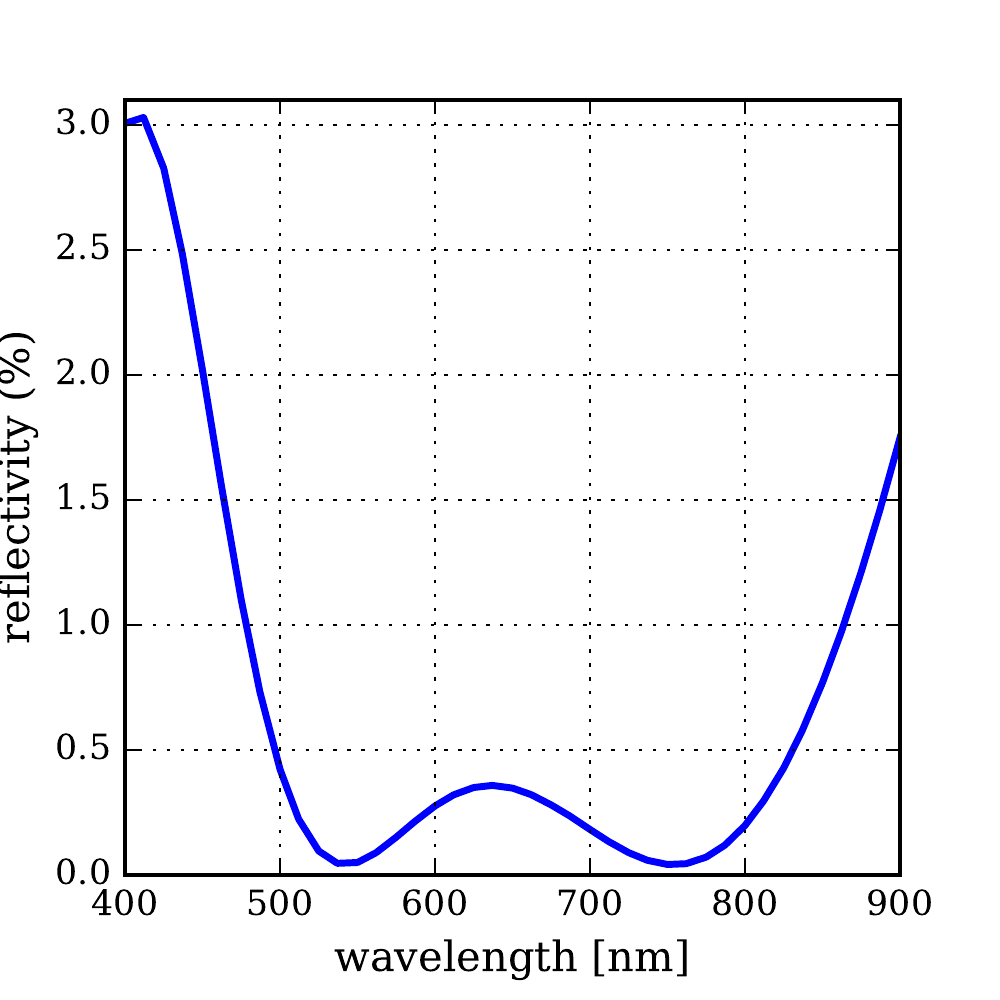}
\caption{{\bf a)}: Optimal geometry of a single bump of the square lattice with \added{$h_{\rm lower} = 225$\,nm, $h_{\rm upper} = 244$\,nm, $w_{\rm bottom} = 244$\,nm, $w_{\rm middle} = 204$\,nm, $w_{\rm top} = 57$\,nm, and $p = 254$\,nm.} \removed{$h_{\rm lower} = 224.6$\,nm, $h_{\rm upper} = 243.8$\,nm, $w_{\rm bottom} = 244.1$\,nm, $w_{\rm middle} = 204.0$\,nm, $w_{\rm top} = 57.4$\,nm, and $p = 254.1$\,nm.}
\removed{$h_{\rm lower} = 227.6$\,nm, $h_{\rm upper} = 236.7$\,nm, $w_{\rm bottom} = 236.4$\,nm, $w_{\rm middle} = 222.6$\,nm, $w_{\rm top} = 66.4$\,nm, and $p = 237.4$\,nm.}
  Note that the periodicity $p$ is only 10\,nm larger than the bottom width $w_{\rm bottom}$ of the bumps. Furthermore, the electric field is visualized by an intensity profile and electric field vectors evincing large field intensity in the gaps between the bumps. 
{\bf b)}: Wavelength scan of the reflectivity for the optimized structure. The metasurface has an average reflectivity 0.19\% in the wavelength range of 500\,nm to 800\,nm and 0.71\% in the range of 400\,nm to 900\,nm.}
\label{fig:optimum_MS}
\end{figure}

The best geometry with the parameter values \removed{$h_{\rm lower} = 224.6$\,nm, $h_{\rm upper} = 243.8$\,nm, $w_{\rm bottom} = 244.1$\,nm, $w_{\rm middle} = 204.0$\,nm, $w_{\rm top} = 57.4$\,nm, and $p = 254.1$\,nm}\added{$h_{\rm lower} = 225$\,nm, $h_{\rm upper} = 244$\,nm, $w_{\rm bottom} = 244$\,nm, $w_{\rm middle} = 204$\,nm, $w_{\rm top} = 57$\,nm, and $p = 254$\,nm} has an objective value of 0.27\% reflectivity. 
In Fig.~\ref{fig:optimum_MS}, the optimal geometry is shown together with a full wavelength scan of its reflectivity in the range of 400\,nm to 900\,nm. As expected, the reflectivity outside the optimization range of 500\,nm to 800\,nm increases significantly. Still, the average reflectivity in the range of 400\,nm to 900\,nm is only 0.71\%.

In a previous numerical study on the same silicon metasurfaces a significantly higher minimal averaged reflectivity of 4\% was reached in the range of 400\,nm to 900\,nm by means of parameter scans within a three-dimensional parameter space (periodicity of square array, width and height of silicon bumps)~\cite{proust2016optimized}. This again demonstrates, that an optimization within a larger parameter space can significantly improve the performance of nano-optical structures with respect to parameter scans in a small subspace.

\section*{Conclusion}
\label{sec:conclusion}

We compared five state-of-the-art optimization methods and applied them to three characteristic nano-optical optimization problems: the maximization of the coupling efficiency of a single-photon source, the reconstruction of geometrical parameters based on scatterometry data, and the suppression of the reflectivity of a silicon metasurface. The optimization methods were extended to meet typical requirements of computational nano-optics, e.g. several objective function values can be computed in parallel. All methods were run with typical numerical settings, i.e. without a manual targeted adaptation to the optimization problems.

The numerical experiments showed that Bayesian optimization reaches good objective function values in only a fraction of the run times of the other considered methods, downhill simplex optimization, L-BFGS-B, particle swarm optimization, and differential evolution. The use of derivative information with respect to some or all parameters can further reduce the run times of Bayesian optimization. The method can require a significant computational overhead for finding parameter samples with a large expected improvement. However, a corresponding prolongation of the total optimization time could be prevented by computing the next sample in advance and by adapting the effort for finding the sample to the effort of computing the objective function.
While this means that the quality of the samples decreases if they are requested at a high frequency, Bayesian optimization offers a significant performance gain over other methods even for frequencies of more than one sample every 20 seconds (single-photon source optimization and parameter reconstruction). Hence, the method is likely to be advantageous for many 2D and almost all 3D simulation problems. We note, however, there can be situations when the considered Bayesian optimization is inefficient. This is, e.g., the case if the underlying Gaussian process regression is inaccurate due to irregularities such as discontinuities or sharp resonance peaks of the objective function~\cite{calandra2016manifold}\removed{ (see also supporting information)}. \added{Another challenge for Bayesian optimization are high-dimensional optimization problems since the underlying problem of maximizing the expected improvement becomes increasingly hard. The supporting information contains a discussion on Gaussian process regression and its limitations as well as a benchmark study of a high-dimensional optimization problem.}

\added{Finally, we note that the}\removed{The} properties of the optimized single-photon source and the anti-reflective metasurface are considerably better than those obtained previously from parameter scans. The results suggest that, whenever possible, one should perform global optimizations within the physically feasible parameter space in order to assess the technological potential of a nano-optical structure.

\section*{Funding Information}
This project has received funding from the European Unions Horizon 2020 research and innovation programme under the Marie Sklodowska-Curie grant agreement No 675745 (MSCA-ITN-EID NOLOSS).
Further, we acknowledge financial support from the EMPIR programme co-financed by the Participating States and from the European Union’s Horizon 2020 research and innovation programme under grant agreement number 17FUN01 (BeCOMe). We also acknowledge support by KIT through the Virtual Materials Design (VIRT-MAT) project by the Helmholtz Association via the Helmholtz program Science and Technology of Nanosystems (STN). 

\begin{suppinfo} 
Further details on the numerical framework, the implementation of the \added{optimization and simulation} algorithms, \added{a discussion on the robustness of the optimization results, a study on high-dimensional optimization,} and a visualization of the different optimization strategies.
\end{suppinfo}

\bibliography{bibl}

\end{document}


\begin{abstract}
This document provides supporting information to "Benchmarking Five Global Optimization Approaches for Nano-optical Shape Optimization and Parameter Reconstruction" regarding the implementation and numerical setting of the optimization and simulation methods. It also contains a discussion on the robustness of the optimal geometries, a study on high-dimensional optimization, as well as a visualization of the different optimization strategies.
\end{abstract}

\section*{Implementation and Settings of Optimization Methods}

In the field of optical simulations one has often access to computing clusters or powerful multicore computers. Many numerical frameworks such as the python package scipy enable only a sequential optimization. That is, only one objective function value is evaluated at a time. In order to allow for a parallel evaluation of the objective function and for a distribution of the computation of the objective function to several machines, we have integrated the considered optimization methods in a server-client framework. The following python code represents a small example for setting up a constrained optimization study on the client side:
\begin{lstlisting}[language=Python]
# Definition of the search domain
domain = [
    {'name': 'x1', 'domain': (-2,2)},
    {'name': 'x2', 'domain': (0,20)}
]

# Definition of constraints, here x1 < x2, 
# i.e. x1 - x2 < 0
constraints = [
    {'name': 'constraint1', 'constraint': 'x1 - x2'}
]

# Setup of the optimization study with 1000 
# iterations and 4 parallel objective evaluations
study = client.create_study(domain,constraints)
study.set_parameters(max_iter=1000, num_parallel=4)

# Optimization run
while not study.is_done():
    # Obtain a new suggestion from the server
    suggestion = study.get_suggestion()
    # Start a new thread to compute the objective
    threading.Thread(target=acquire, args=(suggestion,)).start()

#thread-based parallel acquisition of the objective
def acquire(suggestion):
    # Compute the objective, e.g. on a cluster
    observation = ComputeObjective(**suggestion.kwargs)
    # Report back the observation to the server
    study.add_observation(observation, suggestion.id)
    
\end{lstlisting}

We adapted the optimization methods to support inequality constraints of the parameter space that arise from geometrical or practical (e.g. fabrication) requirements. That is, one may provide a function $f_{\rm cons}: \mathbf{x}\in\mathcal{X} \rightarrow \mathbb{R}$ for each constraint, such that only parameters $\mathbf{x}$ in the search domain $\mathcal{X}\subset \mathbb{R}^D$ with $f_{\rm cons}(\mathbf{x})\leq0$ are sampled. The search domains of the three optimization problems described in the main text are summarized in Tab.~\ref{tab:search_domains}.

Optimization algorithms generally depend on one or more behavioral parameters, such as population sizes and mutation constants, which can strongly influence the performance of the algorithms. However, the meta-optimization of the behavioral parameters for a specific design problem often requires thousands of optimization runs of the underlying design problem~\cite{pedersen2010tuning}. Since for the considered problems a single optimization run takes about 10 hours, a meta-optimization is not practical in our context. Therefore, for every algorithms the standard parameters of the respective optimization packages are chosen, which are generally a reasonable choice for many optimization problems.

\begin{table*}[ht]
\caption{Parameter ranges of the search domains for the three optimization problems. All length scales are given in nanometers,  the angle $\phi_{\rm swa}$ is given in degree, and the scaling factors $s_i$ ($i=1,2,3$) are without units. See the main text for the definition of the parameters.}
\label{tab:search_domains}
\begin{center}
\begin{tabularx}{0.52\linewidth}{lrrrrrr}
\hline 
\multicolumn{7}{c}{Single photon source}\\
\hline 
 & $h_{\rm layer}$ & \added{$r_{\rm core}$}\removed{$d_{\rm core}$} & \added{$r_{\rm mesa}$}\removed{$d_{\rm mesa}$} & $h_{\rm mesa}$ & $h_{\rm dip}$ & $s_{\rm mf}$ \\
min. & 10  & \added{500}\removed{1000} & \added{250}\removed{500}  & 500  & 0    & 0 \\
max. & 150 & \added{1500}\removed{3000} & \added{750}\removed{1500} & 1500 & 1000 & 1000 \\
\hline 
\end{tabularx}
\begin{tabularx}{0.52\linewidth}{lrrrrrrr}
\multicolumn{8}{c}{Parameter reconstruction}\\
\hline 
 & ${\rm CD}$ & $h_{\rm line}$ & $\phi_{\rm swa}$ & $r_{\rm top}$ & $d_{\rm groove}$ & $\sigma_{\rm r}$ & $s_i$ \\
min. & 60 & 115 & 80 & 1  & 1  & 0.1 & 0.1\\
max. & 70 & 125 & 90 & 20 & 25 & 3.0 & 3.0 \\
\hline 
\end{tabularx}
\begin{tabularx}{0.52\linewidth}{lllllll}
\multicolumn{7}{c}{Antireflective metasurface}\\
\hline 
 & $p$ & $h_{\rm lower}$ & $h_{\rm upper}$ & $w_{\rm bottom}$ & $w_{\rm middle}$ & $w_{\rm top}$ \\
min. & 150 & 25  & 25  & 50  & 50  & 10  \\
max. & 300 & 300 & 300 & 290 & 250 & 250 \\
\hline 
\end{tabularx}
\end{center}
\end{table*}

\subsection*{Local Optimization Methods}

For a parallelization of local optimization methods, several independent local optimization runs are started from different points in the parameter space $\mathcal{X} \subset \mathbb{R}^D$. To get a good coverage of the parameter space, we draw starting points from a pseudorandom Sobol sequence~\cite{sobol1967distribution,sobol_py}. After a local optimization has converged, it is restarted at a new point from the Sobol sequence. In all benchmark problems, \removed{the local optimization methods are started from 10 different initial points}\added{the sampling points are drawn alternatingly from 10 local optimizations}. 

For optimization problems that do not exploit derivative information, we consider the downhill simplex algorithm~\cite{doi:10.1093/comjnl/7.4.308}. For a $D$ dimensional parameter space $\mathcal{X}\subset\mathbb{R}^D$, the simplices consist of $D+1$ points $\mathbf{x}_0,\cdots,\mathbf{x}_D$. In each step the point $\mathbf{x}_h$ with the largest function value is replaced by a better one by testing candidates along the line connecting the central point of the simplex and $\mathbf{x}_h$. If this fails, the simplex is contracted and eventually converges to a local minimum. We initialize the simplices such that they span 10\% of the search domain $\mathcal{X}$ in each parameter dimension.

As a gradient-based method we consider the L-BFGS-B algorithm~\cite{byrd1995limited}. This method moves at each step in the direction of the steepest descent. In order to determine the step size, the method constructs a low-memory approximation of the inverse of the Hessian matrix of second partial derivatives. 

If the local optimization methods compute a sampling point $\mathbf{x}$ that does not meet a constraint, i.e. $f_{\rm cons}(\mathbf{x}) > 0$, the sampling point is not sent to the client for evaluation. Instead, the optimization method is provided with the function value $f_{\rm max} + f_{\rm cons}(\mathbf{x})$ and the gradient $\nabla f_{\rm cons}(\mathbf{x})$, where $f_{\rm max}$ is the maximal function value seen so far. This procedure is repeated until the optimization method computes a sampling point that meets the constraints.

Both local optimization methods are implemented based on the python package \texttt{scipy.optimize}~\cite{scipy}.

\subsection*{Stochastic Global Optimization}

As stochastic global optimization methods, we consider particle swarm optimization and differential evolution.
 
Particle swarm optimization has a simple algorithm. Each particle holds the information of the position $\mathbf{p} \in \mathcal{X} \subset \mathbb{R}^D$ of the lowest function value seen so far by the particle, and the position $\mathbf{s}\in\mathcal{X}$ of the lowest function value seen by the swarm. At each step random numbers $r_d$ and $q_d$ are chosen uniformly from $[0,1]$ for each direction $1\leq d \leq D$ and each particle independently. The velocity $v_d$ of a particle in direction $d$ is updated according to
\begin{equation*}
v_d \leftarrow \omega v_d + \varphi_p r_d (p - x_d) + \varphi_{\rm S} q_d (s_d - x_d),
\end{equation*}
where $x_d$ is the current position of the particle in direction $d$. That is, the updated velocity is a weighted sum of the previous velocity, a velocity in direction of $\mathbf{p}$, and in direction of $\mathbf{s}$. The position of particle in direction $d$ is updated according to $x_d \leftarrow x_d + v_d$. 

The implementation is based on the Python package \texttt{pyswarm}~\cite{pyswarm}, which supports a parallel evaluation of the objective function. We extend the algorithm by initializing the particles by a Sobol sequence, excluding parameter values that do not meet the constraints. Furthermore, the random velocity and position update is repeated until the position meets all constraints. If this fails 100 times, the particle is randomly placed somewhere else in the constrained parameter space $\mathcal{X}\subset \mathbb{R}^D$. For the benchmark, we use the standard configuration of \texttt{pyswarm}, i.e. the weights $\omega, \varphi_p, \varphi_{\rm S}$ are set to $0.5$ and the swarm consists of 100 particles. 

Differential evolution is a population-based genetic algorithm. The next generation is computed by creating a new offspring $\mathbf{y}$ for each individual $\mathbf{x}$ in the population by a weighted average (crossover) of some individuals $\mathbf{a},\mathbf{b},\mathbf{c} \in \mathcal{X}$ of the current generation. According to a fixed \emph{crossover probability} (also termed \emph{recombination constant}), random parameter dimensions $0\leq i \leq D$ are selected. For these dimension, the parameter value of the offspring is set to $\mathbf{y}_i = \mathbf{a}_i + F(\mathbf{b}-\mathbf{c}_i)$, where $F$ is called \emph{differential weight} or \emph{mutation constant}. 
If the offspring is fitter than its parent (i.e. it has a lower objective function value), it replaces the parent in the next generation. Differential evolution is implemented based on the python package \texttt{scipy.optimize.differential\_evolution}~
\cite{scipy,scipy.differentialevolution}.  
We use the standard parameters of the optimizer, i.e. \texttt{strategy='best1bin'}, \texttt{mutation=(0.5, 1)}, \texttt{recombination=0.7}. Moreover, we use the default population size of 15 times the number $D$ of parameters. As in the case of particle swarm optimization, the algorithm is extended by initializing the individuals by a Sobol sequence meeting the constraints and by repeating the crossover until the offspring fulfills the constraints. Furthermore, we implemented a thread-based parallel evaluation of the fitness function.

\begin{figure}[ht]
\center
\includegraphics[width=\linewidth]{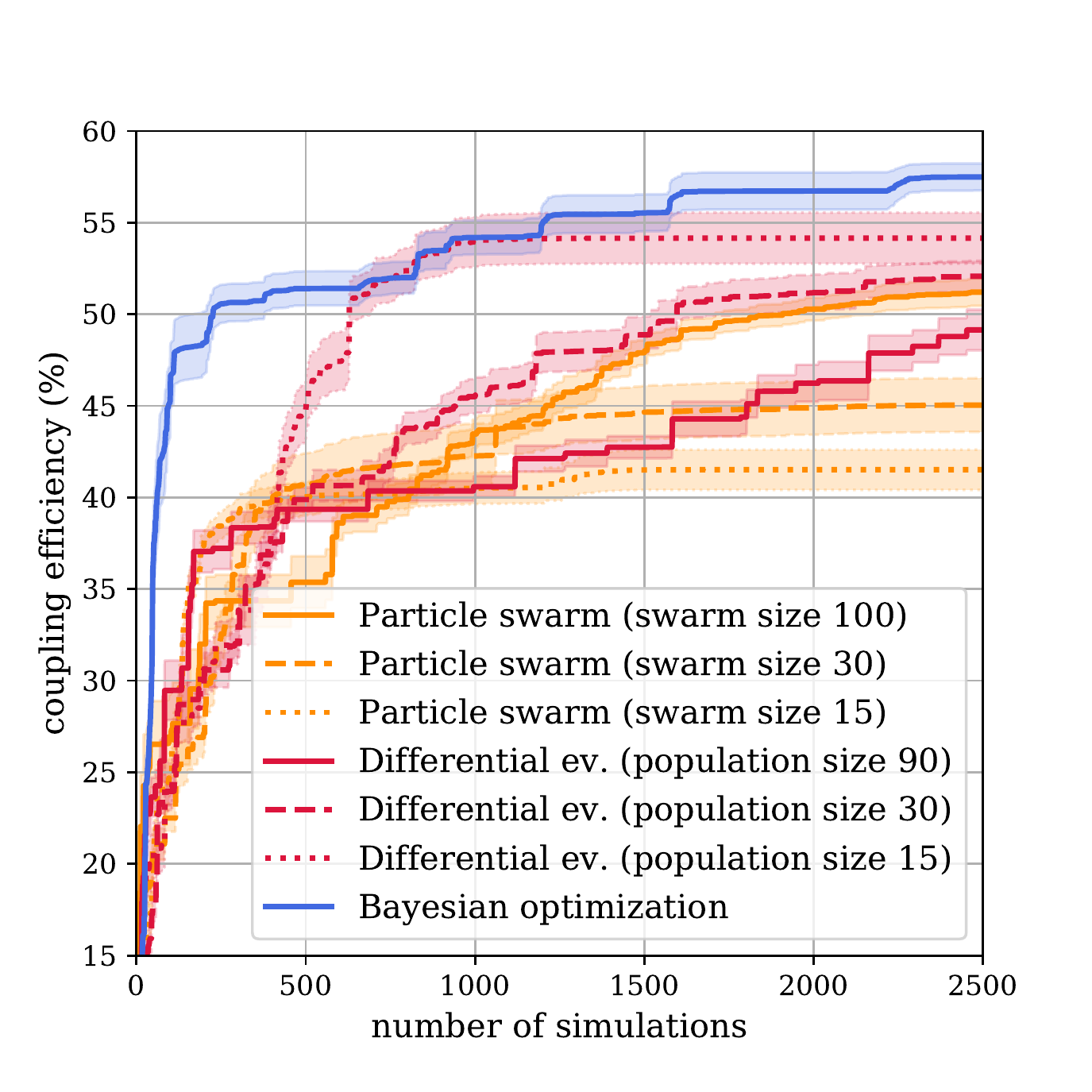}
\caption{ Best seen coupling efficiency for particle swarm optimization and differential evolution for different swarm or population sizes as a function of the number of simulations averaged over six independent optimization runs. The shading indicates the \removed{uncertainty of the average}\added{standard error}. The average performance of Bayesian optimization is also shown for comparison.}
\label{fig:Comp_parameters}
\end{figure}

The performance of the stochastic optimization methods can strongly depend on the chosen optimization parameters. For example, Fig.~\ref{fig:Comp_parameters} shows the average performance of particle swarm optimization and differential evolution for different swarm and population sizes, respectively. Generally, a small size leads to a fast convergence to a local minimum, but to a poor exploration of the parameter space. While the overall performance of particle swarm increases with the swarm size, the opposite holds for differential evolution. We note, that although differential evolution with a \removed{smaller population size} \added{small populations size of 15} is converging  much better, it is still outperformed by Bayesian optimization for all three benchmark problems. 

\subsection*{Bayesian Optimization}

Bayesian optimization is based on a stochastic model of the objective function, which is usually a Gaussian process (GP). Given previous observations of the objective functions, a GP can predict the function value and its statistical uncertainty for each point of the parameter space $\mathbf{x}\in\mathcal{X}$ by means of GP regression~\cite{shahriari2016taking}. 
The accuracy of the prediction can be enhanced by incorporating derivative information on the objective function~\cite{solak2003derivative}. If available, we use this information in order to speed up the optimization process~\cite{wu2017exploiting}.

Based on this statistical information one can determine the expected improvement ${\rm EI}(\mathbf{x})=\mathbb{E}[{\rm max}(0,f_{\rm min} - f(\mathbf{x}))]$, i.e. the probabilistic expectation value of the one-sided difference ${\rm max}(0,f_{\rm min} - f(\mathbf{x}))$ between the function value $f(\mathbf{x})$ and the currently known lowest function value $f_{\rm min}$. The next sampling point is chosen at a position of maximized expected improvement.

In order to parallelize Bayesian optimization, we follow a proposal by Gonzalez {\em et al.}~\cite{pmlr-v51-gonzalez16a}. Through a penalization function $\phi(\mathbf{x})$, regions close to the parameter values of running calculations are avoided. That is, the next sampling point is chosen at a point of maximal penalized expected improvement
\begin{equation*}
\mathbf{x}^* = \underset{\mathbf{x} \in \mathcal{X}}{\operatorname{arg\,max}}\left[\phi(\mathbf{x}) {\rm EI}(\mathbf{x})\right].
\end{equation*}

The search process for the next sampling point is itself an optimization problem that can be computationally demanding~\cite{shahriari2016taking}.
We base the search for $\mathbf{x}^*$ on a differential evolution on the constrained search space followed by an L-BFGS-B optimization of the found maximum. Letting both approaches run with standard parameters until convergence can easily take several minutes. In the considered parallel optimization scenarios this can be far too time consuming. For example, after 2,500 iterations an average search time of 30 seconds would sum up to a total overhead of 20 hours which is longer than the total optimization times of the considered benchmark problems. Therefore, we use two strategies in order to eliminate this time overhead:
\begin{itemize}
\item  After every five iterations, the maximum sample computations time $t_{\rm sample}^{\rm (max)}$ and the the average objective computation time $\overline{t}_{\rm obj}$ is determined. The updated populations size of the differential evolution $N_{\rm pop}$ is chosen such that the sample computation time is restricted to the average computation time $\overline{t}_{\rm comp}$ divided by the number of parallel computations $N_{\rm parallel}$, i.e. $N_{\rm pop}= \overline{t}_{\rm comp}/t_{\rm sample}^{\rm (max)}\cdot N_{\rm pop}^{\rm (prev)}/N_{\rm parallel}$, where $N_{\rm pop}^{\rm (prev)}$ is the previous population size. 
 \item In order to avoid a time delay $\overline{t}_{\rm sample}$ while maximizing the expected improvement, one sampling point is computed in parallel to the $N_{\rm parallel}$ evaluations of the objective function based on the penalization strategy by~\citet{pmlr-v51-gonzalez16a}
\end{itemize}
We note, that this approach still requires that $\overline{t}_{\rm comp}/N_{\rm parallel}$ is much larger than the GP evaluation $t_{\rm GP}$ time. For the example of the single photon source, one has $\overline{t}_{\rm comp}/N_{\rm parallel}\approx17\,s$. The search process after 2000 objective function evaluation was based on average on 1,100 GP evaluations and took about 10 seconds (i.e. $\overline{t}_{\rm comp}/N_{\rm parallel} \approx 2,000 t_{\rm GP}$). This number of GP evaluations was still sufficient to identify and converge into new maxima of the coupling efficiency.

The behavior of the GP is determined by a kernel function $k: (\mathbf{x},\mathbf{x'})\in\mathcal{X}^2 \rightarrow \mathbb{R}$ that defines the covariance between each pair of points in the parameter space $\mathcal{X}\subset \mathbb{R}^D$. We model the {\it a priori} unknown covariance function by a Mat\'{e}rn $5/2$ function
\begin{eqnarray*}
\label{eq:M52}
k_{\rm M52}(\mathbf{x},\mathbf{x}') &=& \sigma^2\left(1 + \sqrt{5} r + \frac{5}{3}r^2 \right)  \exp\left(-\sqrt{5} r\right)\\
\;\;\text{with}\;\; 
r &=& \sqrt{\sum_{i=1}^D (x_i - x_i')^2/l_i^2}.
\end{eqnarray*}
The hyperparameters $\sigma$ and $l_1,l_2,\cdots,l_D$ determine the standard deviation and the length scales of the GP. Having drawn some samples of the objective function, we optimize their values by maximizing the marginal likelihood of the drawn samples with respect to the hyperparameters~\cite{1742-6596-963-1-012003}.

\added{We note that objective functions should be sufficiently smooth to be efficiently modelled by the Mat\'{e}rn $5/2$ kernel.
For example, sharp resonance peaks would lead to extremely short length scales that prevent an interpolation between known data points and an efficient optimization of the objective function. Moreover, there is ongoing research about the question which smoothness class of the objective function is required in the context of Bayesian optimization.
Many results in this direction are based on the strong connection between a GP and the reproducing kernel Hilbert space (RKHS) $\mathcal{H}_{k_\alpha}$ corresponding to the Mat\'{e}rn kernel $k_\alpha$ of the GP (in our case $\alpha=5/2$)~\cite{kanagawa2018gaussian}. For example, Bull~\cite{bull2011convergence} shows that the expected improvement strategy converges at a near-optimal rate of $\mathcal{O}(n^{-1/D})$ for $\alpha > 1$ and for objective functions $f:\mathcal{X}\rightarrow\mathbb{R}^D$ with $f \in \mathcal{H}_{k_\alpha}$. As a consequence, using the Mat\'{e}rn kernel with $\alpha=5/2$ kernel implies that the algorithm converges efficiently if the objective function is $\lfloor\alpha\rfloor = 2$ times differentiable in the classical sense. Moreover, the RKHS $\mathcal{H}_{k_\alpha}$ is norm equivalent to the Sobolev space $W^{\alpha + D/2}_2$ of weakly differentiable functions up to order $5/2 + D/2$. However, this result does not imply, that Bayesian optimization necessarily fails if this smoothness properties are not fulfilled.  Vasquez and Bect~\cite{vazquez2010convergence} show, for example, that for a fixed Gaussian process prior $\pi$, the expected improvement strategy converges almost surely to the minimum for $f$ drawn from the prior $\pi$. The smoothness of a function drawn from the Gaussian is $D/2$ times smaller than the smoothness of the corresponding Kernel (i.e. it lies in $W^{\alpha'}_2$ with $\alpha' + D/2 <\alpha$)~\cite{kanagawa2018gaussian} such that a larger class of objective functions can be optimized. Indeed, it is still an open question whether Bayesian optimization converges for \emph{all} continuous functions~\cite{vazquez2010convergence}. For the case of discontinuities, extended methods, such as manifold GP regression have to be considered~\cite{calandra2016manifold}.}

\removed{We note that only continuous and sufficiently smooth objective functions can be modeled by the Mat\'{e}rn $5/2$ kernel. Features such as discontinuities or sharp resonance peaks would lead to extremely short length scales that prevent an interpolation between known data points and an efficient optimization of the objective function. In this case, extended methods, such as manifold GP regression have to be considered~\cite{calandra2016manifold}.}

\section*{Numerical Simulation Methods}
\revision_remark{Added section}
The behavior of all considered systems 
was determined based of discretization of the structures into higher-order finite elements~\cite{BurgerZschiedrichPomplunetal.2015}. To this end the software package JCMsuite~\cite{JCMsuite} was employed. Important features for the considered systems are an adaptive expansion of rotationally symmetric 3D problems into cylindrical function, the automatic differentiation with respect to input parameters, and an adaptive perfectly matched layer (PML) to realize absorbing boundary conditions.

The coupling efficiency of the single photon source has been determined by computing the propagating modes of the singlemode fiber as well as the scattering field of a dipole source located at the quantum-dot position (see Fig.~1a of the main text). In a second step, the overlap of the scattering field with the fiber mode was computed in order to determine the fraction of the photon energy that is emitted into the fiber~\cite{Schneider:18}.

For the parameter reconstruction of the lamellar grating, the structure is discretized assuming invariance along the grating lines and periodicity perpendicular to the grating lines. This leads to an effective 2D scattering problem~\cite{soltwisch2017reconstructing}.

The metasurface is defined by a square unitcell with periodic boundary conditions in the two lateral directions. A vertically incident plane wave is scattered by the metasurface into different transmitting and reflecting plane waves. The reflectivity is given by the energy flux of the reflected plane waves divided by the energy flux of the incoming plane wave.

The bandpass filter is realized by a layered geometry, which is invariant in two dimensions. This leads to an effective 1D scattering problem, i.~e. the structure is discretized into a 1D grid. The transmittance is again determined by the energy flux going through the layer stack divided by the incoming energy flux.

\section*{Robustness and Sensitivity}
\revision_remark{Added section}

\begin{figure*}[ht]
\center
\raisebox{4.0cm}{a)}\includegraphics[width=0.45\textwidth]{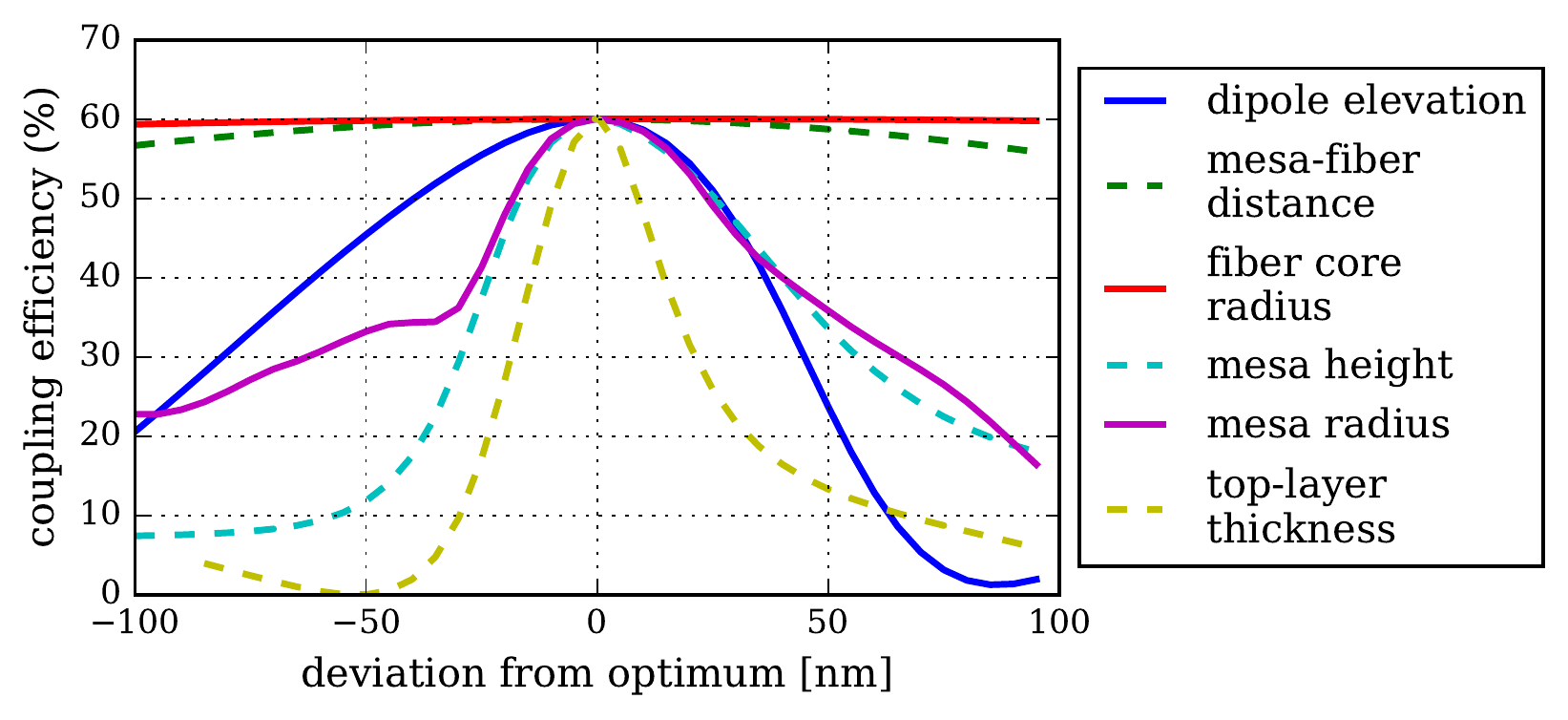}
\raisebox{4.0cm}{b)}\includegraphics[width=0.45\textwidth]{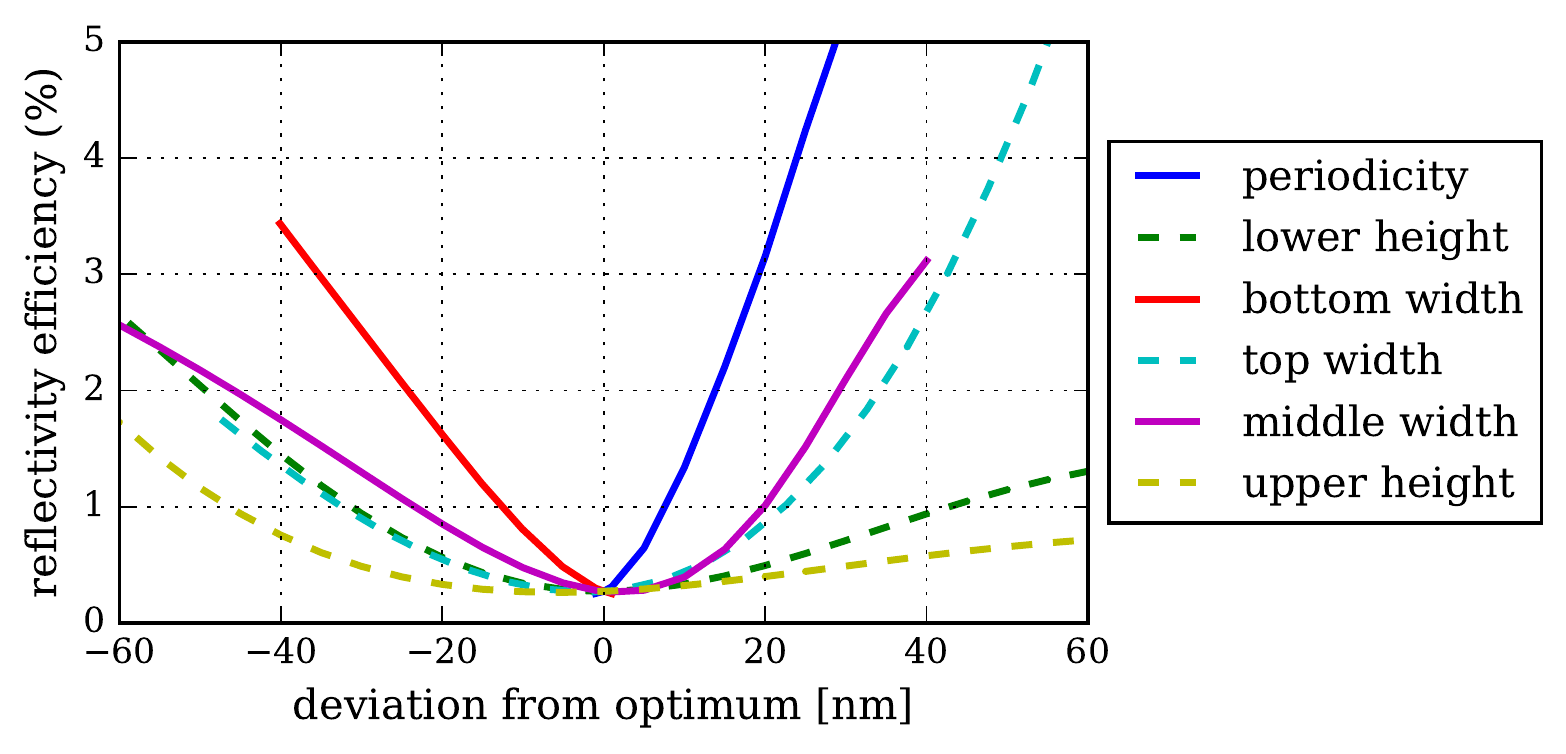}
\caption{
  {\bf a)} Sensitivity of the coupling efficiency of the optimized single photon source to deviations from the optimal values of the system parameters given in the main text.
  {\bf b)} Sensitivity of the reflectivity of the optimized antireflective metasurface to deviations from the optimal values of the system parameters given in the main text. The magnitude of the maximal deviations differ due to the constraints of the parameter space. E.g., the bottom width cannot be larger than the periodicity length of the metasurface. 
}
\label{fig:sensitivity}
\end{figure*}

The numerical optimization of optical devices usually has to incorporate the limitations of the employed fabrication techniques. By applying inequality constraints on the parameter space, one can typically exclude geometries that cannot be fabricated. Another important limitation are fabrication variances, like errors in the targeted geometrical parameters or unintended surface roughnesses~\cite{Schneider:18}.
Figure~\ref{fig:sensitivity} shows the sensitivity of the objective value with respect to deviations from the optimal parameter values for the optimized design of the single photon source and the antireflective metasurface. Clearly, the sensitivity strongly varies from parameter to parameter. The coupling efficiency of the single photon source, for example, is much more sensitive to fluctuations in the top-layer thickness than to the radius of the fiber core. The low reflectivity of the metasurface is very sensitive to periodicity and the bottom width, i.e. the bottom width of the bumps must match as close as possible to the periodicity.

Also, fabrication variances can largely differ between the system parameters~\cite{Schneider:18}. Ideally, the performance of an optimal design should be robust against the expected fabrication variances. That is, a targeted design $\mathbf{x}_0\in\mathcal{X}$ that minimizes the objective $f(\mathbf{x})$ can be perturbed to a value $\mathbf{x}^* = \mathbf{x}+\delta\in\mathcal{X}^*$ with probability distribution $p(\delta;x)$. Hence, a robust average design objective is given as
\begin{equation}
 f_{\rm avg}(\mathbf{x}) = \int_{\mathcal{X}^*} f(\mathbf{x}^*)p(\delta;\mathbf{x}){\rm d}\delta.
\end{equation}
Alternatively, one can define the robust worst-case objective with respect to some probability level $p_0$
\begin{equation}
 f_{\rm max}(\mathbf{x}) = \max_{p(\delta;\mathbf{x})\geq p_0} f(\mathbf{x}^*).
\end{equation}

Unfortunately, the optimization of the robust design objectives is often impractical because the highdimensional integration or maximization of the expensive black-box function $f(\mathbf{x}^*)$, which has to be performed at every optimization step, can be extremely time consuming. A more efficient approach in the context of Bayesian optimization is to perform the integration or maximization over the Gaussian process model of the objective instead~\cite{beland2017bayesian,bogunovic2018adversarially}.

\section*{A High-Dimensional Optimization Problem}
\revision_remark{Added section}

For the three considered medium-dimensional optimization problems, Bayesian optimization shows to be superior in comparison to the other optimization methods. It is an interesting question to which extent this also holds for higher-dimensional problems. 
It is known that high-dimensional optimization is a challenge for Bayesian optimization since the underlying problem of maximizing the expected improvement becomes increasingly hard. Many different methods have been proposed to improve the scalability of Bayesian optimization for large dimensionality, e.g., vie random embeddings~\cite{wang2013bayesian} or by optimizing only a subset of parameters at each step~\cite{li2018high}.

\begin{figure*}[ht]
\center
\raisebox{4.8cm}{a)}\includegraphics[width=0.3\textwidth]{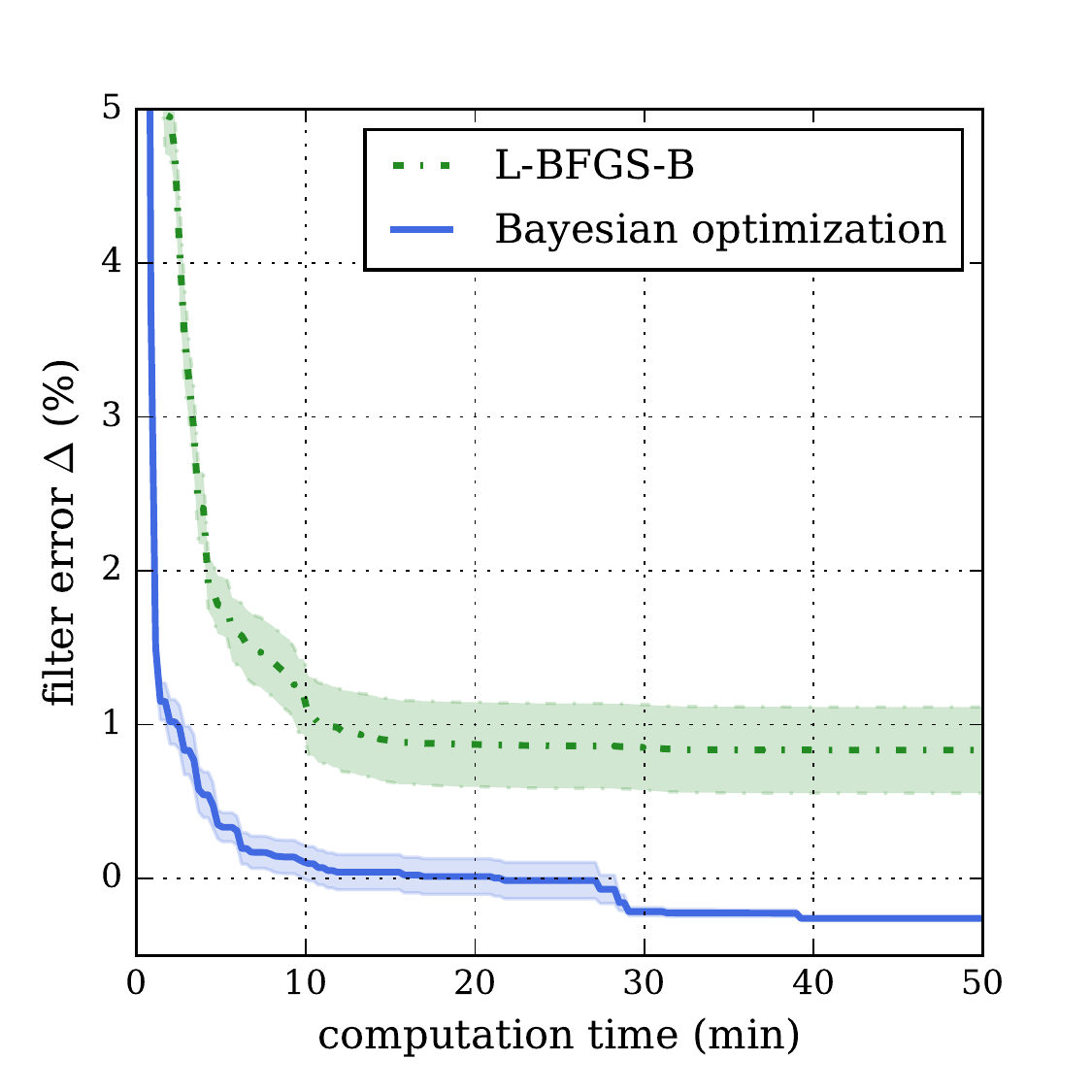}
\raisebox{4.8cm}{b)}\includegraphics[width=0.3\textwidth]{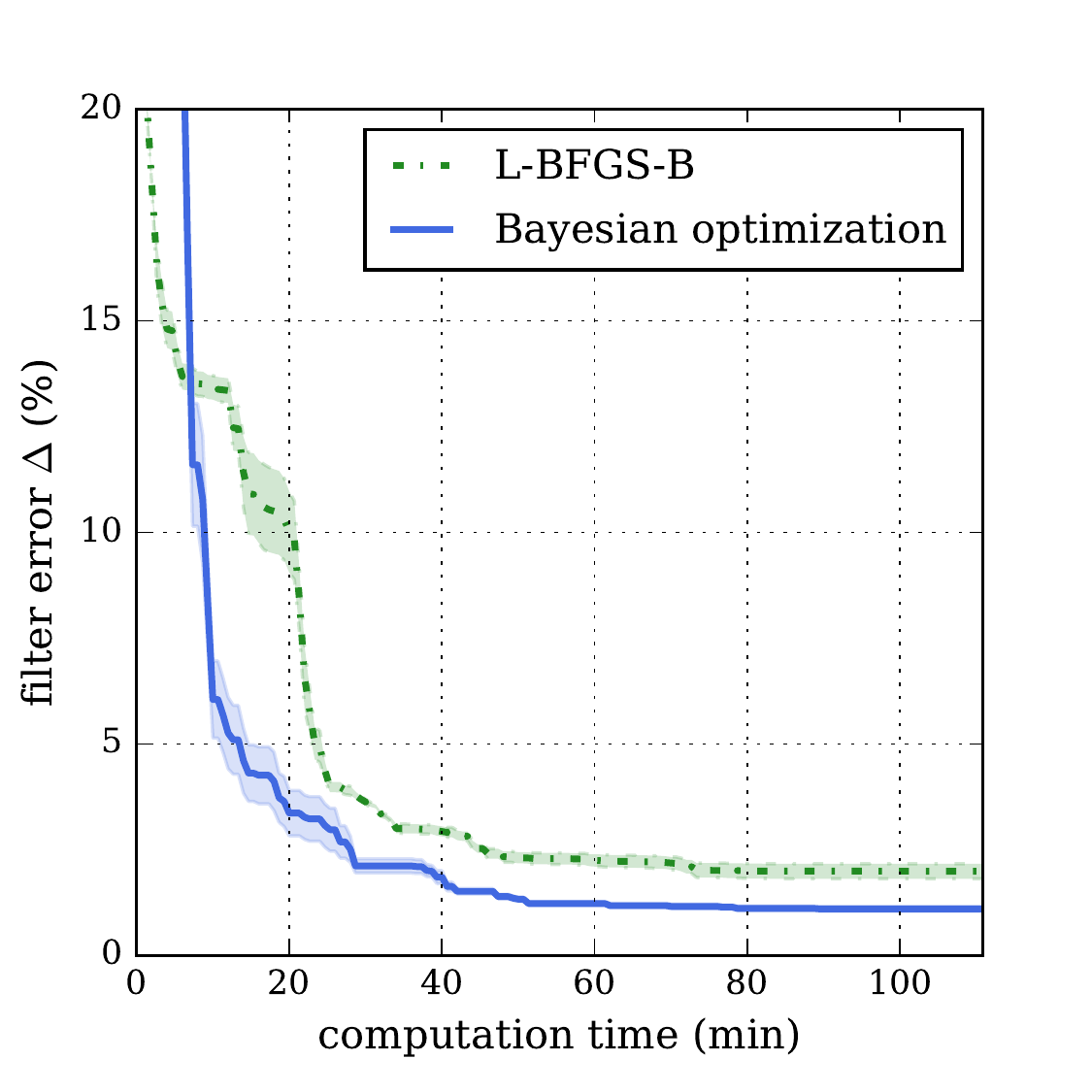}
\raisebox{4.8cm}{c)}\includegraphics[width=0.3\textwidth]{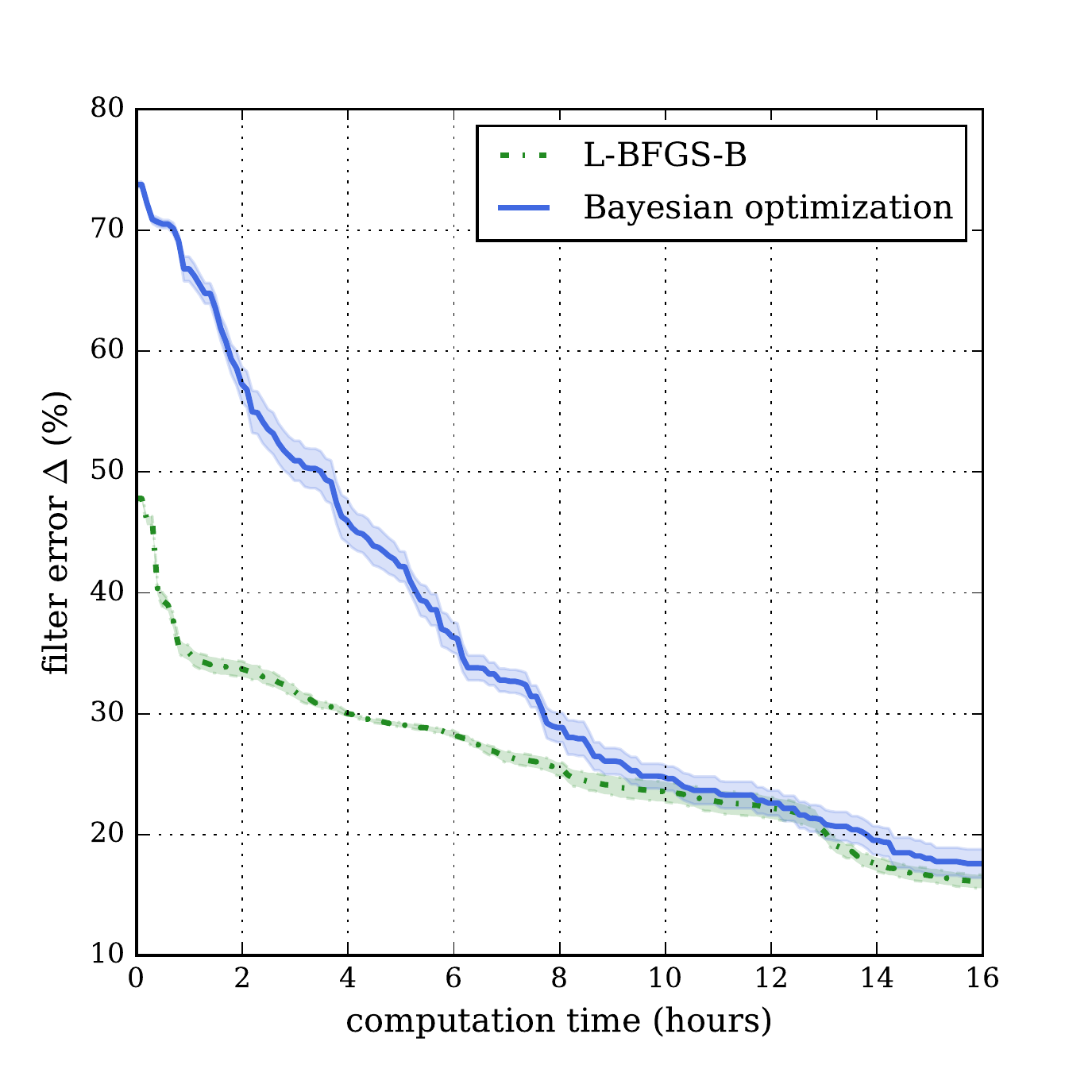}
\caption{
  {\bf a)}~Best seen filter error with 7 DOF ($N=3$ layer pairs) as a function of the total computation time averaged over six independent optimization runs. The shading indicates the standard error of the average.  
  {\bf b)}~The same as in a) but with 21 DOF ($N=10$ layer pairs).
  {\bf c)}~The same as in a) with 51 DOF ($N=25$ layer pairs).
}
\label{fig:high-dim}
\end{figure*}

The following study investigates the performance of the gradient-based Bayesian approach explained above for an increasing number of dimensions. We compare the performance to a local gradient-based optimization with L-BFGS-B. As a benchmark problem, we consider the optimization of a spectral filter consisting of alternating layers of a dielectric layer $D$ made of silicon oxide (SiO$_2$) and a metallic layer $M$ made of silver. The device has the general structure $D(MD)^N$ and has been previously studied Hassan \emph{et al.}~\cite{hassan2015optimal}. 
By varying the thicknesses of the $2N+1$ layers, the objective is to maximize the transmittance in a short-wavelength regime between $\lambda_{\rm short} = 850\,$nm and $\lambda_{\rm gap} = 1,780\,$nm and to minimize the transmittance in a long-wavelength regime between $\lambda_{\rm gap}$ and $\lambda_{\rm long}=6,500\,$nm. Given the average transmission $T_{\rm pass}$ in the short-wavelength regime  and the average transmission $T_{\rm stop}$ in the long-wavelength regime, the error of the filter is defined as
\begin{equation}
  \Delta = \max(\tau - T_{\rm pass},T_{\rm stop}-\beta),
\end{equation}
where $\tau=79\,\%$ defines the targeted minimal transmittance for short wavelengths and $\beta=2\,\%$ the targeted maximal transmittance for long wavelengths~\cite{hassan2015optimal}.

We run the optimization for devices with $N\in\{3,10,25\}$ layer pairs corresponding to problems with 7, 21, and 51 degrees of freedom (DOF), respectively. 
The dielectric layers are varied between 100\,nm and 400\,nm and the metallic layers between 2\,nm and 5\,nm.
The initial layer thicknesses for the local optimization are set to 4\,nm for the metallic layer and 242\,nm for the dielectric layers, i.e. a quarter-wave thickness with respect to the gap wavelength $\lambda_{\rm gap}$~\cite{hassan2015optimal}.
In order to determine the average performance of the local optimization method, the initial parameters are randomly disturbed within a range of 10\% of the search space between each optimization run.
We note, that the purpose of increasing the number of layers is merely to define problems with increasing DOF. More layers do not necessarily lead to a better optimal device performance since a larger number of metallic layers reduces the transmittance also in the short-wavelength regime.

Figure~\ref{fig:high-dim} shows the average convergence of Bayesian optimization and L-BFGS-B. Both optimization methods run sequentially for 100 iterations, i.e.~only samples from a single L-BFGS-B optimizer are taken and Bayesian optimization does not compute samples while a calculation is running. For $D=7$ DOF, Bayesian optimization clearly converges faster to small filter errors than L-BFGS-B. For a growing number of DOFs, these roles are exchanged. Bayesian optimization needs more and more time for computing new samples since after $M$ iterations, the underlying covariance matrix has $[M(D+1)]^2$ entries, which slows down the inversion of the matrix and the calculation of the expected improvement. For example, for $D=51$ the computation of a new sample takes after 50 iterations about half a minute. Moreover, the global strategy leads to an increasing overhead for a large number of dimensions. Altogether, gradient descent methods like L-BFGS-B are therefore usually more suited for high-dimensional optimization problems. \newpage

\section*{Optimization Strategies}

\begin{figure*}[ht]
\centering
\includegraphics[width=0.83\textwidth]{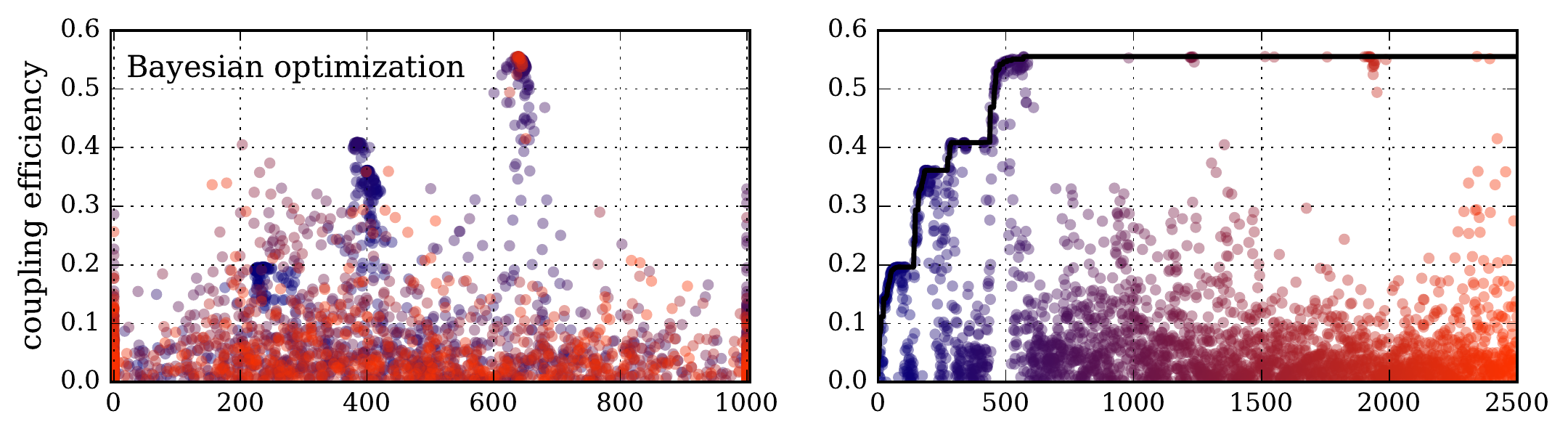}
\includegraphics[width=0.83\textwidth]{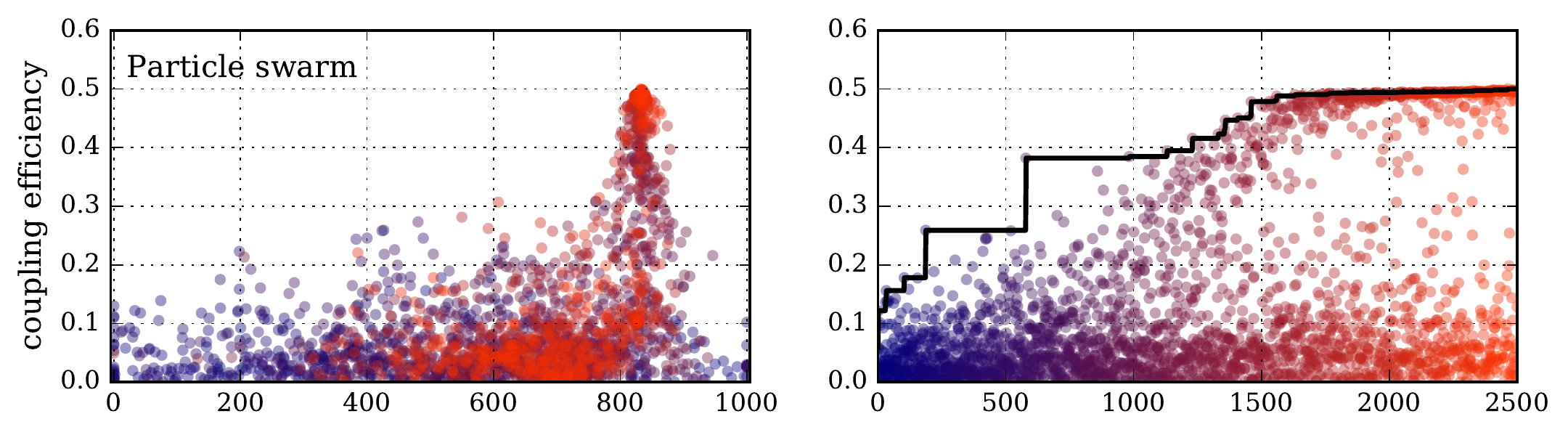}
\includegraphics[width=0.83\textwidth]{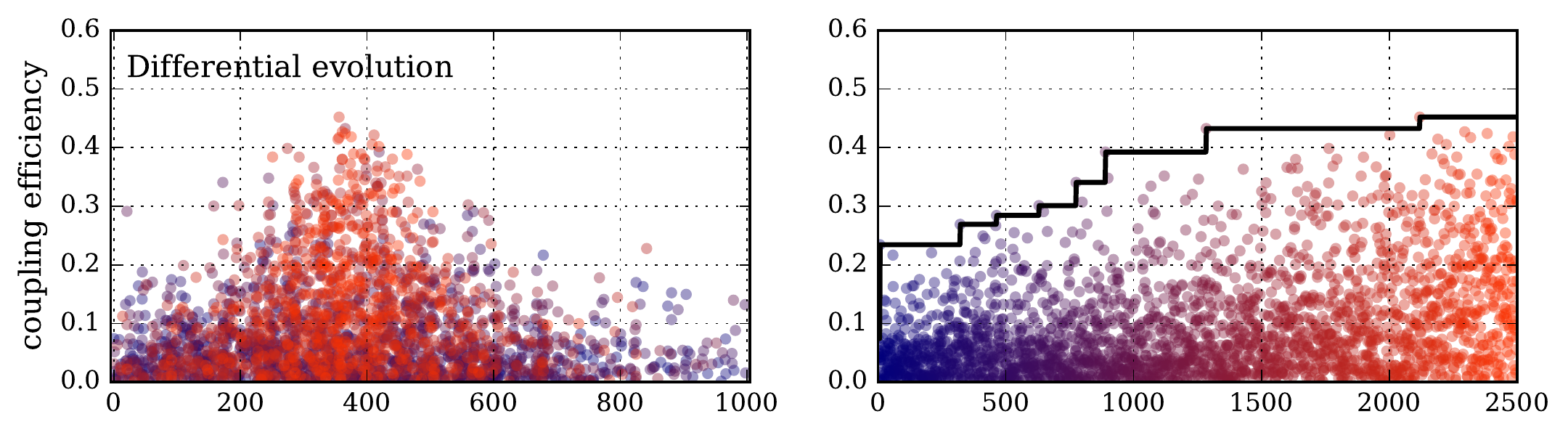}
\includegraphics[width=0.83\textwidth]{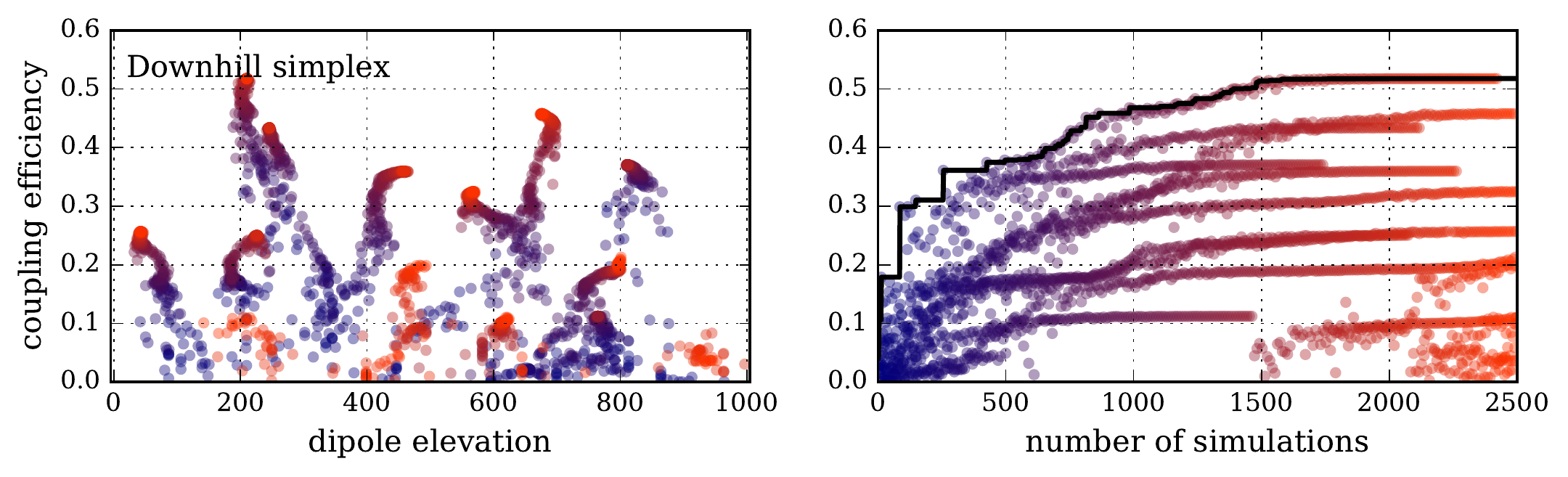}
\caption{{\bf Left column}: Objective function value plotted against  dipole elevation $h_{\rm dip}$ of the single photon source for all samples drawn during a run of the different optimization approaches.
{\bf Right column}: Best seen objective function value plotted against the number of simulations for the same optimization run.
The points are colored according to the number of simulation (dark blue at the beginning, light red at the end of the optimization).
The graphs visualize the different strategies of the optimization approaches. 
{\bf Bayesian optimization} probes sequentially different local minima. Whenever the expected improvement within a local minimum gets small, other parts of the parameter space are sampled until a better minimum is found (exploration). If this fails, the best found minimum is probed again (exploitation).
{\bf Particle swarm optimization} first probes the parameter space randomly. After about 500 simulations a good sample with close to 40\% coupling efficiency is found. A part of the swarm moves into the direction of this swarm minimum.
{\bf Differential evolution} builds new population members by changing some parameter values of existing members (point mutation). This leads to the formation of vertical lines on the left side (i.e. different individuals with the same dipole elevation). 
{\bf Downhill simplex optimization} performs parallel local optimizations starting from different initial points. Each optimization run converges to a different local optimum. After convergence, the local optimization is restarted from a new position.
}
\label{fig:visualization}
\end{figure*}

The strategies of the different methods are visualized in Fig.~\ref{fig:visualization} for the example of the single photon source optimization. For all methods the achieved coupling efficiencies are plotted as a function of one geometry parameter and as a function of the optimization progress. The figure shows, for example, that Bayesian optimization alternates between phases of exploitation (convergence into a found local minimum) and exploration (sampling away from all known local minima). Figure~\ref{fig:visualization} reveals that the objective function has a large number of local minima since each parallel run of the downhill simplex algorithm converges to a different minimum.

\clearpage
\bibliography{bibl}